\documentclass[prl, nofootinbib, notitlepage,twocolumn]{revtex4-1} 

\usepackage{graphicx}
\usepackage{caption}
\usepackage{subcaption}
\usepackage{amsmath,bm}
\usepackage[T1]{fontenc}
\usepackage{rotating}
\usepackage{float}
\floatstyle{plaintop}
\restylefloat{table}
\usepackage{soul}

\usepackage{color}
\definecolor{naviBlue}{RGB}{0,0,128}
\usepackage[colorlinks  = true,
            linkcolor   = naviBlue,
            urlcolor    = naviBlue,
            citecolor   = naviBlue,
            anchorcolor = naviBlue]{hyperref}
\newcommand{\secref}[1]{\hyperref[sec::#1]{SECTION~\ref*{sec::#1}}}
\newcommand{\subsecref}[1]{\hyperref[subsec::#1]{SECTION.~\ref*{subsec::#1}}}
\newcommand{\figref}[1]{\hyperref[fig::#1]{FIG.$\,$\ref*{fig::#1}}}
\newcommand{\tabref}[1]{\hyperref[tab::#1]{TABLE$\,$\ref*{tab::#1}}}
\newcommand{\eqnref}[1]{\hyperref[eqn::#1]{Eq.$\,$(\ref*{eqn::#1})}}

\newcommand{\diff}{\mathrm{d}}
\newcommand{\I}[1]{\textit{#1}}
\newcommand{\sv}{\langle \sigma v \rangle}

\usepackage{ulem}

\begin{document}

\title{Novel dark matter constraints from antiprotons in the light of AMS-02}

\author{Alessandro Cuoco}
\email{cuoco@physik.rwth-aachen.de}
\author{Michael Kr\"amer}
\email{mkraemer@physik.rwth-aachen.de}
\author{Michael Korsmeier}
\email{korsmeier@physik.rwth-aachen.de}
\affiliation{Institute for Theoretical Particle Physics and Cosmology, RWTH Aachen University, 52056 Aachen, Germany}


\begin{abstract}
We evaluate dark matter (DM) limits from cosmic-ray antiproton observations using the recent precise AMS-02 measurements.
We properly 
take into account  cosmic-ray propagation uncertainties, fitting DM and propagation
parameters at the same time, and marginalizing over the latter.
We find a significant  ($\sim$\,4.5 $\sigma$) indication of a DM signal for DM masses near $80$ GeV, 
with a hadronic annihilation cross-section close to the thermal value, $\left\langle \sigma v \right\rangle \approx 3 \times 10^{-26}$~cm$^3$s$^{-1}$.
Intriguingly, this signal is com\-pa\-ti\-ble with the DM interpretation
of the Galactic center gamma-ray excess.
Confirmation of the signal will require a more accurate study of the systematic uncertainties, i.e.,  the antiproton production cross-section, 
and the modeling of the effect of solar modulation.
Interpreting the AMS-02 data in terms of upper limits on hadronic DM annihilation, we obtain strong constraints 
excluding a thermal annihilation cross-section for DM masses below about 50\,GeV and in the range between approximately 150 and 500\,GeV, even
for conservative propagation scenarios.
Except for the range around $\sim\,$80 GeV, our limits are a factor $\sim$\,4 stronger than the limits from gamma-ray observations of dwarf galaxies.
\end{abstract}

\maketitle

\section*{\label{sec::introduction}Introduction}

Cosmic-ray (CR) antiprotons are  a powerful tool to investigate the particle nature of dark matter (DM), 
see, for example, \cite{Bergstrom:1999jc,Donato:2003xg,Bringmann:2006im,Donato:2008jk,Fornengo:2013xda,Hooper:2014ysa,Pettorino:2014sua,Boudaud:2014qra,Cembranos:2014wza,Cirelli:2014lwa,Bringmann:2014lpa,Giesen:2015ufa,Evoli:2015vaa}. DM constraints from CRs are, however, affected by uncertainties in the description of CR propagation in the Galaxy. 
Thus, CR DM limits have so far been derived for benchmark
propagation models, like the MIN/MED/MAX scenarios \cite{Donato:2003xg} obtained from observations of the Boron over Carbon (B/C) ratio. Such benchmark models introduce an order-of-magnitude uncertainty in the DM interpretation of CR fluxes.

The antiproton CR spectrum has recently been measured by the AMS-02 experiment with high precision \cite{Aguilar:2016kjl}.
It is thus timely to evaluate the antiproton DM constraints in the light of the new data. We will improve on previous analyses in two crucial aspects:
First, the new AMS-02 data allow us to significantly reduce the uncertainties in the CR
propagation. Although B/C data from AMS-02 have been recently published~\cite{Aguilar:2016vqr}, there is, however,  evidence that the propagation of heavy nuclei like B and C is different from the propagation
of light nuclei like $p$ and $\bar{p}$ \cite{Johannesson_CR_Propagation_2016} 
(but, see also \cite{Yuan:2017ozr,Jin:2017iwg,Lin:2016ezz}). Thus, using B/C
data to constrain CR propagation is likely to introduce a bias when analysing antiprotons. We will instead follow the analysis of Ref.~\cite{Korsmeier:2016kha} (hereafter KC16)
and use the measured $\bar{p}$ flux to directly constrain the propagation scenario, thus avoiding any bias.
In addition, as a second important new feature, we will constrain CR propagation including a potential $\bar{p}$ flux from DM
annihilation. Previous analyses have, in contrast, assumed a certain propagation scenario (or a small number of fixed benchmark scenarios) and
thus a fixed antiproton background to then constrain a DM contribution in a second step (although, see \cite{Bringmann:2014lpa} for an improved approach).
Here, with a joint DM and CR propagation analysis, we will,  for the first time, explore possible correlations and degeneracies between the two components, providing
more robust and reliable DM constraints.

\section*{\label{sec::DM}Dark Matter}

Dark matter annihilation in the Galaxy leads to a flux of antiprotons from the fragmentation of Standard Model (SM) particles. The corresponding source term can be 
written as:
\begin{eqnarray}
  \label{eqn::DM_source_term}
  q_{\bar{p}}^{(\mathrm{DM})}(\bm{x}, E_\mathrm{kin}) = 
  \frac{1}{2} \left( \frac{\rho(\bm{x})}{m_\mathrm{DM}}\right)^2  \sum_f \left\langle \sigma v \right\rangle_f \frac{\diff N^f_{\bar{p}}}{\diff E_\mathrm{kin}} ,
\end{eqnarray}
where $m_\mathrm{DM}$ is  the  DM mass and $\rho(\bm{x})$ the DM density profile. 
Furthermore, $\left\langle \sigma v \right\rangle_f$ denotes the thermally averaged annihilation cross-section for the SM final state $f$,  
${\rm DM}\!+\!{\rm DM} \to f\!+\!\bar{f}$, and $\diff N^f_{\bar{p}}/\diff E_\mathrm{kin}$ the corresponding antiproton energy spectrum per DM annihilation. 
Note that the factor $1/2$ corresponds to Majorana fermion DM.

We use the NFW DM density profile~\cite{Navarro:1995iw}, 
 $   \rho_{\mathrm{NFW}}(r) = \rho_h \, r_h/r\, \left( 1 + r/r_h \right)^{-2}$, 
with a characteristic halo radius $r_h=20\,$kpc, and a characteristic halo density $\rho_h$, normalized so that to obtain a 
local DM density $\rho_\odot = 0.43\,$GeV/cm$^3$~\cite{Salucci:2010qr} at the 
solar position $r_ \odot = 8\,$kpc. 
To quantify the impact of the choice of the DM profile on our results, 
we will compare with the Burkert profile~\cite{Burkert:1995yz}, 
$    \rho_{\mathrm{Bur}}(r) =  \rho_c \, (1+r/r_c)^{-1} (1 + r^2/r_c^2 )^{-1}$, 
with a core radius of $r_c=5\,$kpc,  and again normalized at the solar position. 

The yield of antiprotons per DM annihilation, and the corresponding energy distribution, ${\diff N^f_{\bar{p}}}/{\diff E_\mathrm{kin}}$, depend on the DM mass, the relevant SM annihilation channels, and on the antiproton yield from fragmentation of SM particles. We employ the results presented in \cite{Cirelli:2010xx}, and focus on the annihilation into bottom quarks, ${\rm DM\; DM} \to b\bar{b}$, for illustration.

\section{\label{sec::methods}Analysis}

To derive predictions for the fluxes of protons, helium and antiprotons near Earth, 
we solve the standard diffusion equation \cite{StrongMoskalenko_CR_rewview_2007} 
using \textsc{Galprop} \cite{Strong:1998fr,Strong:2015zva}.
We assume a cylindrical symmetry for our Galaxy, with a radial extension of $20\,$kpc. The propagation parameters which determine the shape of the injection spectrum 
include 
the spectral indices  of the protons and the heavier species, $\gamma_{1,p},\gamma_{2,p}$ and $\gamma_{1},\gamma_{2}$, respectively, the two break positions, $R_0$, $R_1$,  as well as 
smoothing factors, $s$, $s_1$.  The propagation is assumed to be homogenous and isotropic. It is constrained by the normalisation, $D_0$, and slope, $\delta$, of the diffusion coefficient,  
the velocity of Alfven magnetic waves, $v_A$, connected to reacceleration, the convection velocity, $v_{0c}$, the normalization of the proton and helium fluxes, $A_\mathrm{p}$ and $A_\mathrm{He}$, respectively, the Galaxy's half-height, $z_h$,  and the solar modulation potential, $\phi_\mathrm{AMS}$, 
in the framework of the force-field  approximation.
See the supplemental material and KC16 for more details.  
We also take into account the production of tertiary antiprotons \cite{Moskalenko:2001ya}.
The DM component of the CR flux, finally, is determined by the DM mass, $m_{\rm DM}$, and the DM annihilation cross-section $\sv$, for any given choice of the DM profile and the SM annihilation channel.  

We stress that this is a simplified scenario since diffusion is likely to be non-homogenous and anisotropic at some level.
On the other hand, this simplified model has been able to explain the observations so far and has been   assumed in past studies.
It is thus important to address the implication of the new data within this model.
A critical assessment of this base scenario will be the subject of future studies. 
Nonetheless, for the purposes of this analysis, we can consider the homogenous propagation coefficient  $D_0$ as an
effective parameter describing an average propagation, 
since we propagate only light nuclei, which share similar propagation properties.
Violations of homogeneity will will be manifested as a different $D_0$ for heavier nuclei, which have a different propagation length. 
We discuss this issue in more detail in the supplemental material in relation to Boron and Carbon.

The  above propagation and dark matter parameters are determined in a global fit of the AMS-02 proton and helium fluxes \cite{Aguilar_AMS_Proton_2015, Aguilar_AMS_Helium_2015}, and the AMS-02 antiproton to proton ratio \cite{Aguilar:2016kjl}, complemented 
by proton and helium data from CREAM~\cite{Yoon_CREAM_CR_ProtonHelium_2011, Yoon_CREAM_CR_ProtonHelium_2011} and VOYAGER~\cite{Stone_VOYAGER_CR_LIS_FLUX_2013, Stone_VOYAGER_CR_LIS_FLUX_2013}. The CREAM data extend to large rigidities of up to $\approx 100$~TV, and allow us to determine the position, $R_1$, amount, $\Delta\gamma = \gamma_3-\gamma_2$ and smoothness, $s_1$, of the second break in the rigidity dependence of the source. The VOYAGER data at rigidities of ${\cal O}({\rm GV})$, on the other hand, are used to constrain the solar modulation potential $\phi_\mathrm{AMS}$. 

The ranges of variation of the parameters that enter our prediction of the CR flux are listed in \tabref{Parameters2} next to the fit results. Having fixed the strength and position of the second break in rigidity as in KC16, leaves 16 free parameters 
to be determined from a global fit to the AMS-02, CREAM, and VOYAGER data. 
We use \textsc{MultiNest} \cite{Feroz_MultiNest_2008} to scan this parameter space and derive the corresponding profile likelihoods. 
Details of the global fit are presented in KC16.

\begin{figure*}[t!]
	\centering
	  \includegraphics[width=8cm] {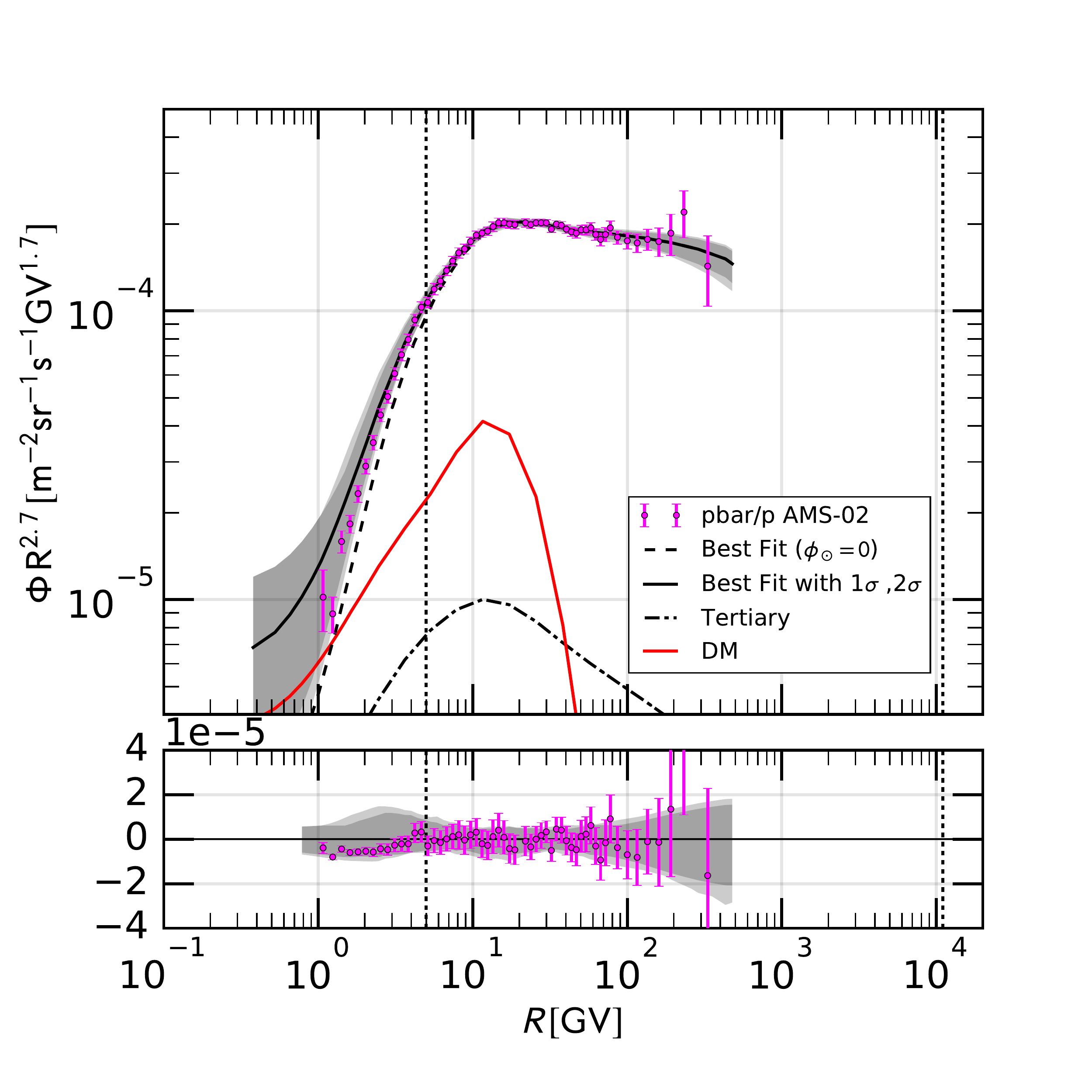}
    \includegraphics[width=8cm] {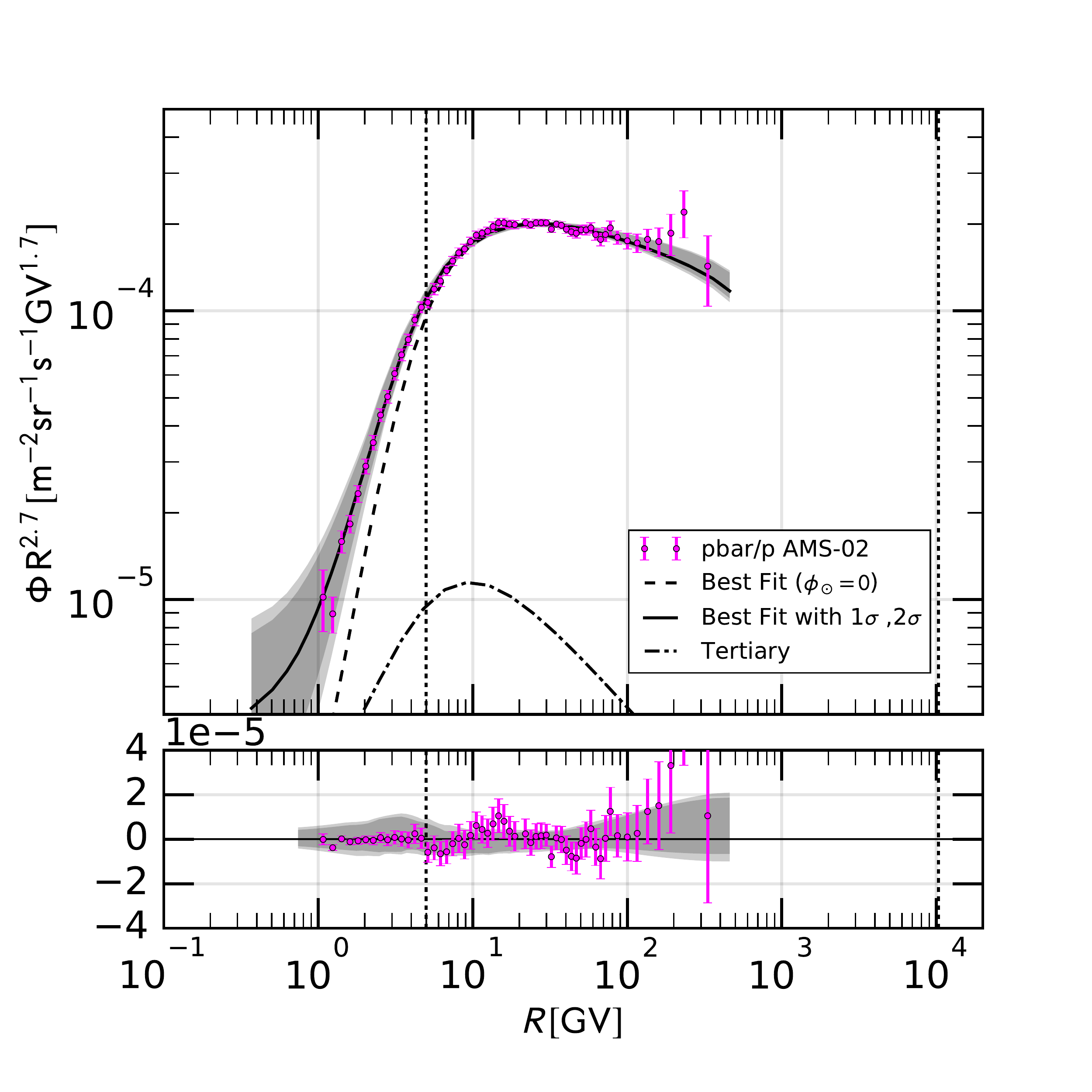}
   	\caption{Comparison of the best fit of the $\bar{p}/p$ ratio to the AMS-02 data~\cite{Aguilar:2016kjl}, with a DM component (left panel) and without DM (right panel). The lower panels show the corresponding residuals. The fit is performed between the dotted lines, \textit{i.e.}, for rigidities $5\,{\rm GV}\le R \le 10\,{\rm TV}$. The grey bands around the best fit indicate the 1 and 2$\sigma$ uncertainty, respectively. 
The dashed black line (labeled ``$\phi_\odot=0$ MV'') shows the best fit without correction for solar modulation.	
The solid red line shows the best fit DM contribution. We also show, for comparison, the contribution from astrophysical tertiary antiprotons denoted by the dot-dashed line.} 
	  \label{fig::pbaroverp_fit}
\end{figure*}

\section{\label{sec::results}Results}

The result of our global fit is shown in \figref{pbaroverp_fit} for the antiproton to proton ratio, for both the case in which DM is included (left panel) and 
the case without a DM component (right panel). 
We consider the rigidity range R $\ge 5$~GV, for which the force-field approximation should describe solar modulation reliably. 
Adding DM annihilating into $b\bar{b}$, 
with $m_{\rm DM} \approx 80$~GeV and $\left\langle \sigma v \right\rangle \approx 3 \times 10^{-26}$~cm$^3$/s, results in a much better fit and provides an intriguing 
hint for a DM signal in the antiproton flux. The improvement of the fit quality is significant: we find a $\chi^2$/(number of degrees of freedom) of 71/165 for the fit without DM, which is reduced to 46/163 when adding a DM component. Formally, $\Delta\chi^2=25$ for the two extra parameters introduced by the DM component corresponds to a significance of $\sim$\,4.5\,$\sigma$,
although this does not take into account possible systematics errors.

The comparison of the two panels provides a deeper insight into the reason for the large improvement of the fit when DM is included.
We can see that, without DM, the residuals show a sharp feature, similar to a break, at a rigidity of  $\approx 18$~GV.
This feature is  present in the measured spectrum and cannot be described  by the
secondary antiprotons only, since their predicted spectrum is too smooth compared to the data. 
We see instead that the DM component, shown separately in the left panel, possesses a distinctive feature which matches  the structure of the residuals
without DM. For comparison, we also show the contribution from background tertiaries, which peaks at similar rigidities,
but which cannot fit the strength and shape of the excess.

The preferred range of DM masses and annihilation cross-sections is shown in \figref{xsection}. 
Intriguingly, this region is in very good agreement with the DM interpretation of the Galactic center gamma-ray excess 
\cite{Gordon:2013vta,Abazajian:2014fta,Daylan:2014rsa,TheFermi-LAT:2015kwa,Calore:2014nla}. We show for comparison the preferred DM best fit region obtained from the Galactic center gamma-ray excess in \cite{Calore:2014nla}. 
Also, a similar hint for DM has been found in \cite{Hooper:2014ysa}, in relation to PAMELA
antiproton data \cite{Adriani:2010rc}.

\begin{figure}[b!]
  \includegraphics[width=0.45\textwidth]{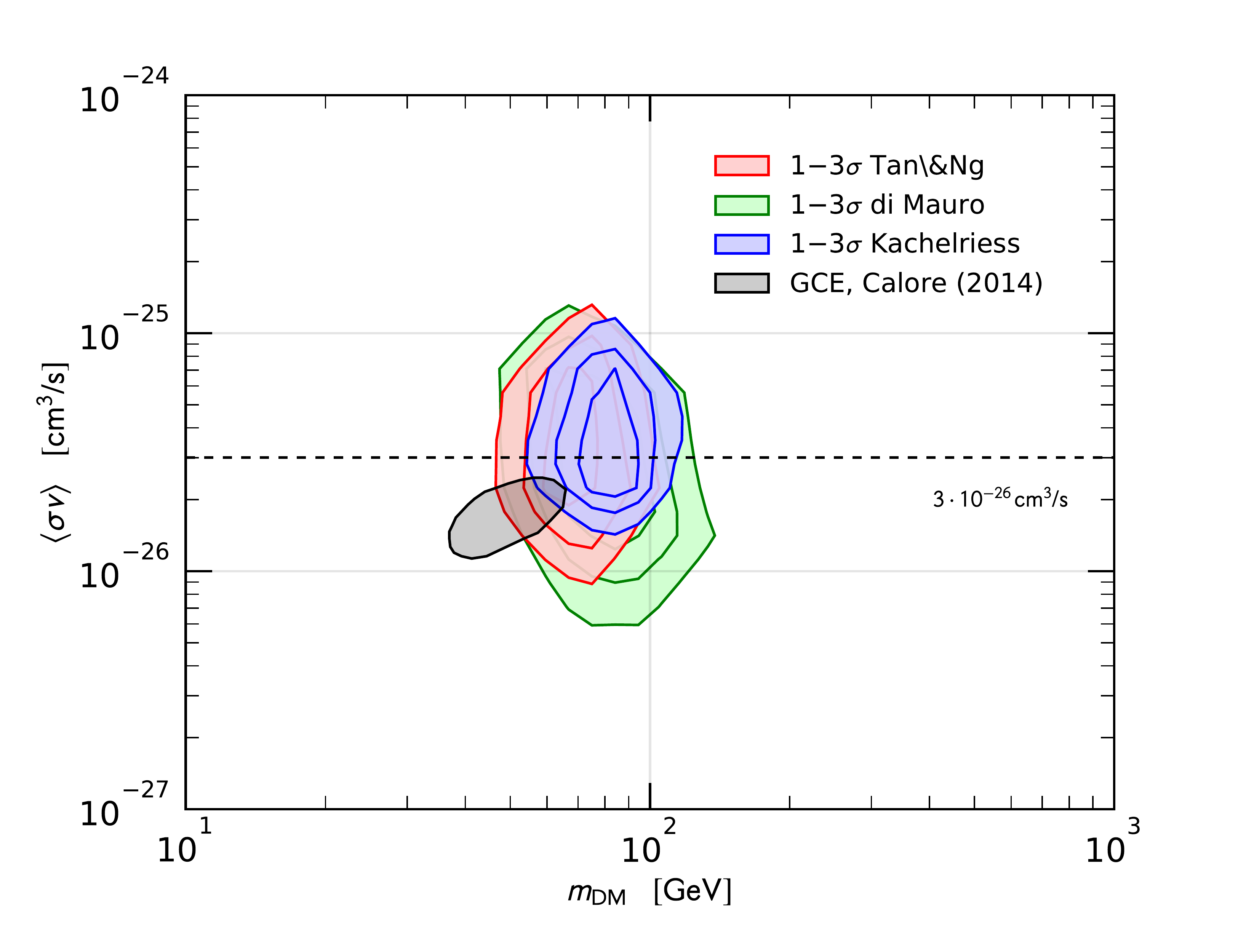}
\vspace{-0.2cm}	    
  \caption{Best fit regions (1, 2 and 3\,$\sigma$) for a DM component of the antiproton flux, using the antiproton cross-section models of \cite{TanNg_AntiprotonParametrization_1983} (Tan \& Ng), \cite{Mauro_Antiproton_Cross_Section_2014} (di Mauro et al.), and \cite{Kachelriess:2015wpa} (Kachelriess et al.). For comparison, we also show the best fit region of the DM interpretation of the  Galactic center gamma-ray excess~\cite{Calore:2014nla}, and the thermal value of the annihilation cross-section, $\left\langle \sigma v \right\rangle \approx 3 \times 10^{-26}$~cm$^3$s$^{-1}$.} 
  	\label{fig::xsection}
\end{figure}

A known systematic uncertainty affecting the fit is the imperfect knowledge of 
the antiproton production cross-section~\cite{Donato:2001ms,Evoli:2011id,Mauro_Antiproton_Cross_Section_2014,Kappl:2014hha,Kachelriess:2015wpa,Winkler:2017xor}, 
which determines the flux of secondary antiprotons produced by the interactions of primary protons and Helium nuclei on the inter-stellar medium gas. 
Adopting the recent cross-section estimates from \cite{Mauro_Antiproton_Cross_Section_2014} and \cite{Kachelriess:2015wpa}, 
rather than the \textsc{Galprop}  default~\cite{TanNg_AntiprotonParametrization_1983},  does not 
reduce the evidence for a DM matter component in the antiproton flux, and modifies only slightly the preferred ranges of DM mass and annihilation cross-section, see~\figref{xsection}.
This represents an important test, since the cross-sections used are quite different in nature.
While those of  \cite{TanNg_AntiprotonParametrization_1983,Mauro_Antiproton_Cross_Section_2014} are based
on a phenomenological parameterization of the available cross-section data, the cross section of  \cite{Kachelriess:2015wpa}
is based on a physical model implemented through Monte Carlo generators.
While this check does not exhaust the range of possible sys\-te\-ma\-tics related to the antiproton cross-section,
a more robust assessment of this issue requires more accurate and comprehensive experimental antiproton cross-section measurements.

\begin{table}[t]
  \caption{Analysis constraints on the fit  parameters,   and their ranges of variation in the fit.}
  \label{tab::Parameters2}
  \centering
  \begin{tabular}{l l l l l }
  \hline \hline
  \multicolumn{2}{l}{ \textbf{ Propagation }}  & \textbf{Fit with-} \hspace{0.1cm}  & \textbf{Standard fit} \\
  \multicolumn{2}{l}{ \textbf{ parameters  }}  & \textbf{out DM   }  & \textbf{with DM} & \textbf{Fit range} \\ \hline \hline
  $\gamma_{1,p}       $  & $                 $ & $1.54   ^{ +0.04       } _{ -0.18      }$ 
                                                & $1.41   ^{ +0.19       } _{ -0.01      }$    
                                                & $1.2$ - $1.8    $ \\
  $\gamma_{2,p}       $  & $                  $ & $2.425  ^{ +0.023      } _{ -0.002     }$
                                                & $2.531  ^{ +0.008      } _{ -0.010     }$     
                                                & $2.3$ - $2.6    $ \\
  $\gamma_1           $  & $                  $ & $1.56   ^{ +0.03       } _{ -0.18      }$
                                                & $1.21   ^{ +0.22       } _{ -0.02      }$     
                                                & $1.2$ - $1.8 $ \\
  $\gamma_2           $  & $                  $ & $2.388  ^{ +0.021      } _{ -0.003     }$
                                                & $2.480  ^{ +0.005      } _{ -0.005     }$     
                                                & $2.3$ - $2.6    $ \\
  $R_0                $  & $[GV]              $ & $8.43   ^{ +0.27       } _{ -1.93      }$
                                                & $5.01   ^{ +1.30       } _{ -0.12      }$     
                                                & $1.0$ - $10     $ \\ 
  $s                  $  & $                  $ & $0.38   ^{ +0.11       } _{ -0.01      }$
                                                & $0.46   ^{ +0.01       } _{ -0.06      }$     
                                                & $0.05$ - $0.9    $ \\
  $\delta             $  & $                  $ & $0.361  ^{ +0.005      } _{ -0.043     }$
                                                & $0.245  ^{ +0.015      } _{ -0.007     }$     
                                                & $0.2$ - $0.5    $ \\
  $D_0                $  & [$10^{28}$ cm$^2$/s] & $7.48   ^{ +\I{1.52}   } _{ -1.88      }$
                                                & $9.84   ^{ +\I{0.26}   } _{ -2.85      }$     
                                                & $0.5$ - $10.0$ \\ 
  $v_\mathrm{A}       $  & [km/s] $           $ & $23.8   ^{ +3.09       } _{ -0.91      }$
                                                & $28.5   ^{ +\I{1.5}    } _{ -0.64      }$     
                                                & $0$ - $30     $ \\ 
  $v_{0,\mathrm{c}}   $  & [km/s] $           $ & $26.9   ^{ +34.7       } _{ -3.33      }$
                                                & $45.3   ^{ +5.69       } _{ -19.2      }$     
                                                & $0$ - $100    $ \\ 
  $z_h                $  & [kpc]  $           $ & $6.78   ^{ +\I{0.22}   } _{ -2.70      }$
                                                & $5.35   ^{ +\I{1.65}   } _{ -1.27      }$     
                                                & $2  $ - $7$  \\
  $\phi_\mathrm{AMS}  $  & [GV]   $           $ & $580    ^{ +65         } _{ -50        }$
                                                & $520    ^{ +35         } _{ -35        }$
                                                & $0 $ - $1.8 $     \\ \hline \hline  
 \multicolumn{2}{l}{ \textbf{DM parameters}}  &   \\ \hline \hline
 \multicolumn{2}{l}{$\log(m_\mathrm{DM}/\mathrm{GeV})$ } & $  $
                                                & $1.85   ^{ +0.02       } _{ -0.03      }$
                                                & $1 $ - $5      $     \\ 
 \multicolumn{2}{l}{$\log(\sv/\mathrm{cm^3/s})       $ } & $  $
                                                & $-25.57 ^{ +0.09       } _{ -0.03      }$
                                                & $-(28$ - $23)    $      \\ \hline \hline 
  \multicolumn{2}{l}{ \textbf{Experiment}}  & \multicolumn{3}{l}{ \textbf{$\chi^2$ (Number of data points)} } \\ \hline \hline
  \multicolumn{2}{l}{$p$ (AMS-02)          }& 9.6 (61)  & 6.2 (61)  \\ 
  \multicolumn{2}{l}{$p$ (VOYAGER)         }& 1.8 (4)   & 0.4 (4)   \\ 
  \multicolumn{2}{l}{He (AMS-02)           }& 30.8 (65) & 24.8 (65) \\ 
  \multicolumn{2}{l}{He (VOYAGER)          }& 2.3 (4)   & 1.6 (4)   \\ 
  \multicolumn{2}{l}{$\bar{p}/p$ (AMS-02)  }& 26.6 (42) & 12.6 (42) \\ \hline 
  Total                 && 71.0 (176) & 45.6 (176) \\ \hline \hline
\end{tabular}
\end{table}


From \tabref{Parameters2}  we note that including a DM component induces a shift in some of the propagation parameters. 
In particular  the slope of the diffusion coefficient, $\delta$, changes by about 30\% 
from a value of $\delta \approx 0.36$ without DM to $\delta \approx 0.25$ when DM is included. 
This stresses the importance of fitting at the same time DM and CR background.
The changes induced by a DM component in the other CR propagation parameters are less than about 10\%.
More details are reported in the supplementary material.

As a further estimate of systematic uncertainties, we have extended the fit range down to a rigidity of $R = 1\,$GV.
In this case, the fit excludes a significant DM component in the antiproton flux. 
This  can be understood from the residuals for this case, which are very similar to the ones shown in the right panel of \figref{pbaroverp_fit}.
Clearly, the excess feature at $R \approx 18\,$GV, responsible for the DM preference in the default case, still remains.
The reason why DM cannot accommodate anymore this excess, is the low-rigidity tail of the DM spectrum,
 \textit{c.f.} \figref{pbaroverp_fit} (left panel), which would overshoot the experimental data below 5 GV.
Nonetheless, although the data at $R \lesssim 5\,$GV appear to disfavor a DM component in the antiproton flux, the situation 
is not conclusive: at rigidities $R \lesssim 5\,$GV, solar modulation deviates from the simple force-field approximation
and exhibits also charge dependent effects \cite{Cholis_Solar_Modulation_2016,Corti:2015bqi}.
Thus, a deeper scrutiny of the antiproton excess and of a potential DM signal will require a dedicated study of the solar 
modulation below 5\,GV, for which it would be desirable to have time series of the proton and antiproton fluxes.

In the remainder of this paper, we will make the conservative assumption of no DM detection and derive  
constraints on the hadronic DM annihilation cross-section as a function of the DM mass.
Our limits on the annihilation cross-section $\left\langle \sigma v \right\rangle$ as a function of $m_{\rm DM}$ are obtained by marginalizing 
over the CR propagation uncertainties. Technically, 
we divide the likelihood samples in the $\left\langle \sigma v \right\rangle$-$m_{\rm DM}$ plane obtained from the
\textsc{MultiNest} scan into 20 slices in $m_{\rm DM}$, equally spaced in $\log(m_{\rm DM})$ between $m_{\rm DM} = 10$~GeV and 100~TeV. 
For each $m_{\rm DM}$ slice we derive the 1D profile likelihood as a function of $\sv$, determining the minimum $\chi^2$
and then set 95\% exclusion limits on $\sv$ from the condition $\Delta \chi^2=3.84$. 
Formally, the correct procedure would amount to fixing  $m_{\rm DM}$ to a grid of values and to perform a separate fit for each of these values. 
However, such a procedure would be computationally very demanding and would lead to results very similar to those 
obtained using the 1D profile likelihood for slices in $m_{\rm DM}$. This is shown in 
 \figref{DM_limits_withSystematics_b}, comparing the black line with the the three black dots,
which are the limits derived with the formally accurate procedure for the three values of $m_{\rm DM}$. 

In order to obtain  an estimate of the systematic uncertainties affecting the limits, we perform fits with
different diffusion models, rigidity cuts, DM profiles, and antiproton production cross-sections. 
The various limits are shown in  \figref{DM_limits_withSystematics_b}.     
Not surprisingly, the worst limits are obtained when fixing the diffusion zone height $z_h$ to the minimal
considered value of 2\,kpc, since in this case a large fraction of a potential DM signal outside 
the diffusion zone cannot reach Earth. Correspondingly, setting the diffusion zone height to the maximal value 
we consider, $z_h=7$\,kpc, leads to a larger DM contribution and thus stronger constraints. Neglecting 
convection in the diffusion equation and/or changing the DM profile from NFW to 
Burkert does not have a significant impact on the fit. 

\begin{figure}[t!]
		  \includegraphics[width=0.45\textwidth]{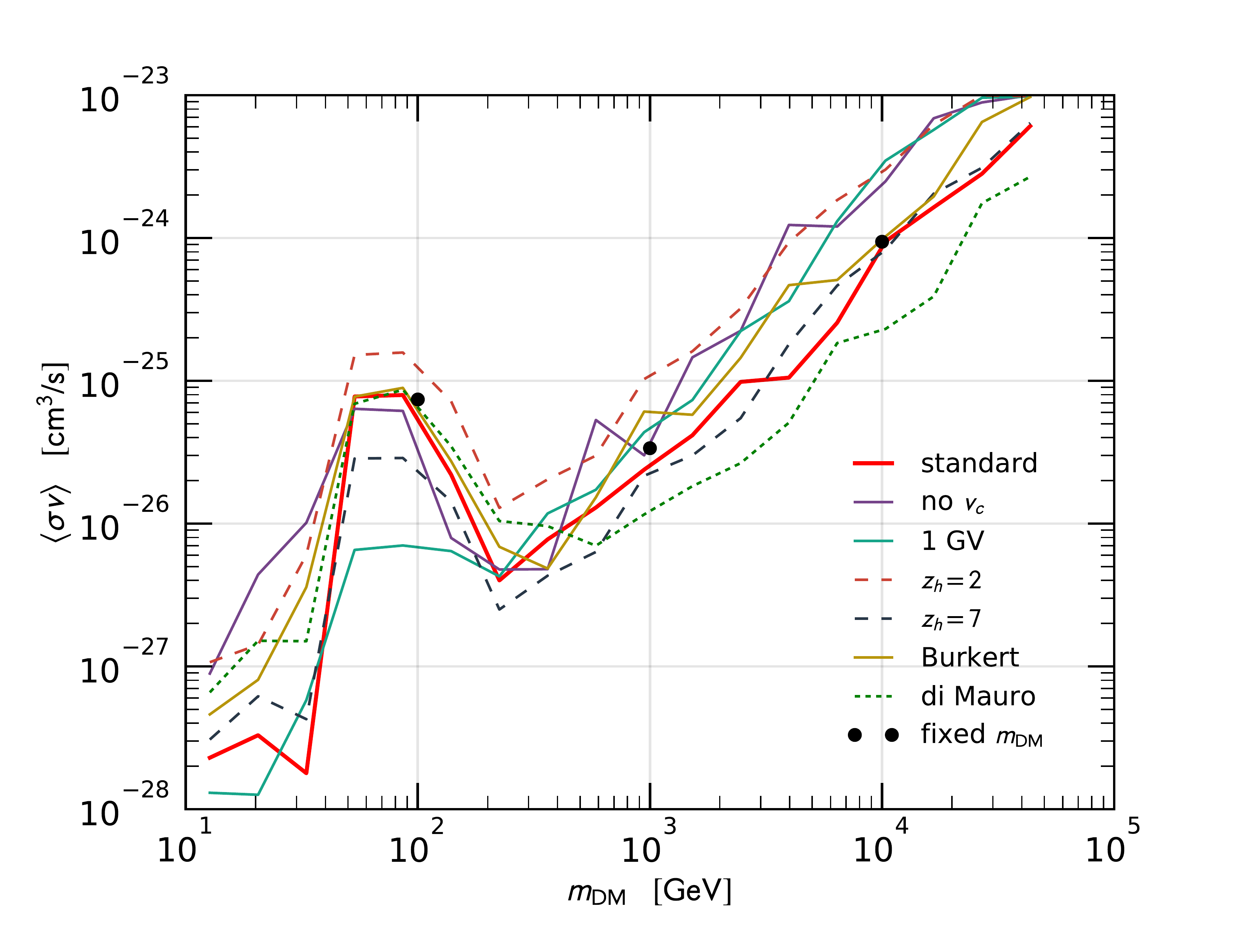}
		  \vspace{-0.2cm}
	    	\caption{Limits on the DM annihilation cross-section into $b\bar{b}$ final states for our standard setting, for different diffusion zone heights, $z_h$, for propagation without convection, for an alternative  
		antiproton cross-section model, for the Burkert DM profile, and for rigidities down to 1\;GV, respectively. We also show limits for three fixed DM masses, as discussed in the text. }
	  \label{fig::DM_limits_withSystematics_b}
\end{figure}

The most prominent feature in  \figref{DM_limits_withSystematics_b} is the weak exclusion near DM masses of 80\,GeV, 
where the fit prefers a significant DM component. The exclusion becomes much stronger for a fit down 
to low rigidities of 1\,GV, which also disfavours a DM signal. However, as argued above, 
the simple force-field approximation is not expected to describe well solar modulation 
at rigidities $R \lesssim 5\,$GV, and more work is needed to interpret the low rigidity data in a reliable way.

We have emphasized the importance of the antiproton production cross-section for a reliable estimate of the 
antiproton flux. Adopting the more recent cross-section model from \cite{Mauro_Antiproton_Cross_Section_2014}, rather than \
the \textsc{Galprop}  default~\cite{TanNg_AntiprotonParametrization_1983}, has little impact on the fit near 
$m_{\rm DM} \approx 80$\,GeV, but the different energy dependence of the cross-section models leads to a change 
in the DM limits for light and heavy DM. 

In \figref{DM_simplelimits} we summarize the result of our fit and show both the evidence for a DM component 
in the CR antiproton flux, as well as limits on the DM annihilation cross-section. The systematic uncertainty on the 
exclusion limit is shown as an uncertainty band obtained from the envelope of the various fits presented in \figref{DM_limits_withSystematics_b}. 
In our baseline scenario (solid line), we can exclude thermal DM with $\left\langle \sigma v \right\rangle \approx 3 \times 10^{-26}$~cm$^3$s$^{-1}$ annihilating into $b\bar{b}$
for DM masses below about 50\,GeV and in the range between approximately 150 and 1500\,GeV. Even considering our most conservative 
propagation scenario, we achieve strong limits and can exclude thermal DM below about 50\,GeV and in the range between approximately 150 and 500\,GeV. 
The results for other hadronic annihilation channels, and for annihilation into $ZZ$ and $W^+W^-$ final states are very similar; in the supplementary material we provide 
limits for DM annihilation in into $W^+W^-$ as a further explicit example. 

In comparison with the results derived in \cite{Ackermann:2015zua} from gamma-ray observations of nearby dwarf galaxies, 
we improve the annihilation cross-section limits by a factor of $\sim$\,4  for all DM masses except those around 80\,GeV.
We also see from \figref{DM_simplelimits} that, similarly to the DM interpretation of the Galactic center gamma-ray excess, 
the preferred region of a DM signal in the antiproton flux is in tension with the dwarf galaxy constraints. However, this tension can be relieved 
with a more conservative estimate of the DM content of the dwarf galaxies~\cite{Bonnivard:2015xpq}.
Also, a recent analysis using new discovered dwarfs galaxies  \cite{Fermi-LAT:2016uux} actually provides weaker limits,
also shown in \figref{DM_simplelimits}, further relieving the tension.

\begin{figure}[t!]
	  \includegraphics[width=0.45\textwidth]{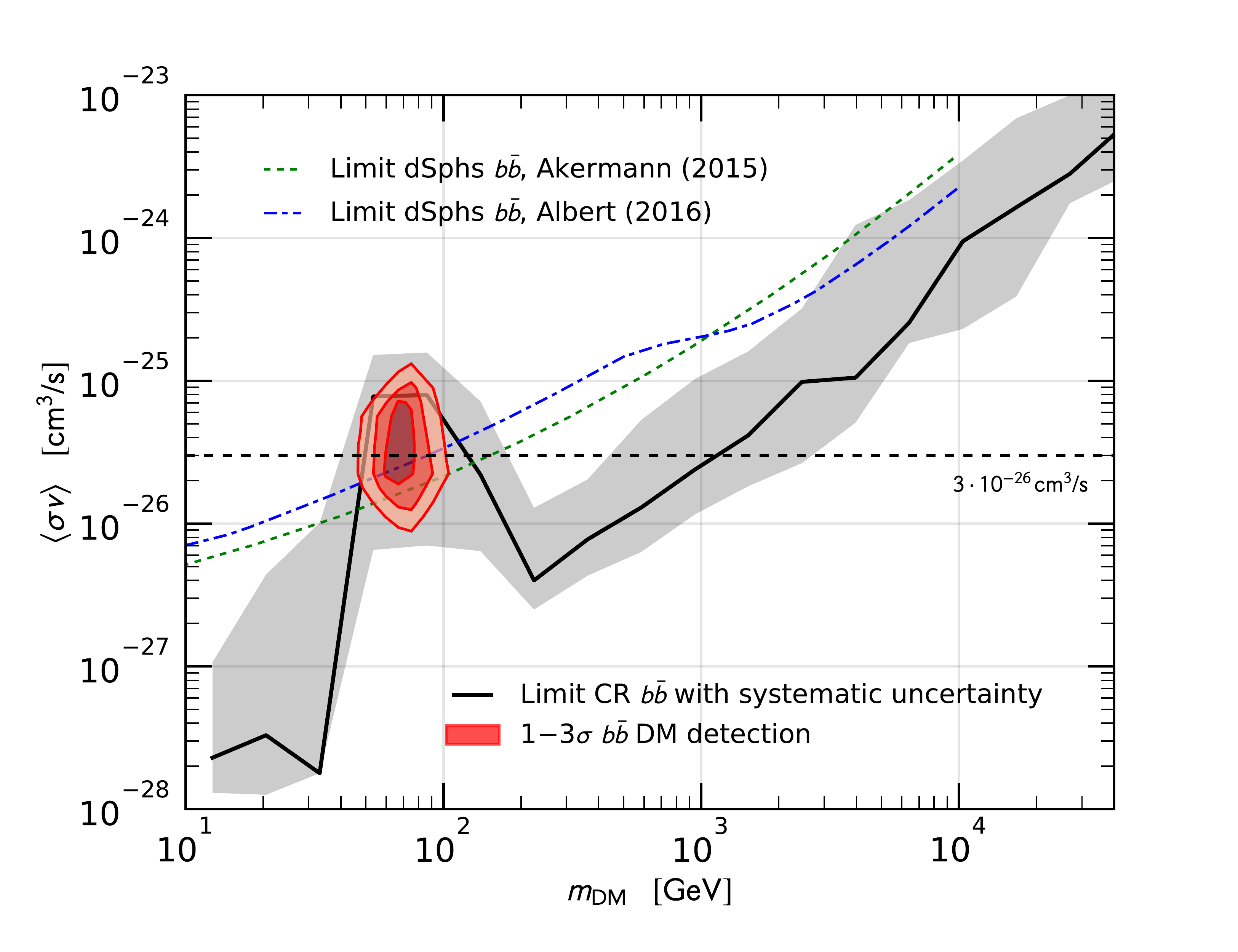}
	  \vspace{-0.2cm}
    \caption{Best fit regions (1, 2 and 3\,$\sigma$) for a DM component of the antiproton flux, and limits on the DM annihilation cross-section into $b\bar{b}$ final states. The grey shaded uncertainty band is obtained from the envelope of the various fits presented in \figref{DM_limits_withSystematics_b}. For comparison we show limits on the annihilation cross-section obtained from gamma-ray observations of dwarf galaxies\,\cite{Ackermann:2015zua,Fermi-LAT:2016uux}, and the thermal value of the annihilation cross-section, $\left\langle \sigma v \right\rangle \approx 3 \times 10^{-26}$~cm$^3$s$^{-1}$.} 
    \vspace{3.9em}
	  \label{fig::DM_simplelimits}
\end{figure}

\section{\label{sec::conclusion}Summary and conclusion}
In conclusion, the very accurate recent measurement of the CR antiproton flux by the AMS-02 experiment allows to achieve
unprecedented sensitivity to possible DM signals, a factor $\sim$\,4 stronger than the limits from gamma-ray observations of dwarf galaxies.

Further, we find an intriguing indication for a DM signal in the antiproton flux, 
compatible with the DM interpretation of the Galactic center gamma-ray excess. A deeper examination of such a potential signal would require 
a more accurate determination of the antiproton production cross-section, to constrain the flux of secondary antiprotons, as well as an accurate modeling 
of solar modulation at low rigidities of less than about 5\,GV.

{\it Note added:} After our submission we became aware of a similar work by \cite{Cui_2016}. They perform an analysis using
methodologies analogous to the ones of this letter  and find results consistent with ours.

\section*{\label{sec::acknowledgments}Acknowledgments}

We wish to thank  Jan Heisig, Julien Lesgourgues, Stefan Schael, and Pasquale Serpico for helpful discussions and comments. 

\bibliography{bibliography}{}

\begin{thebibliography}{10}%
\makeatletter
\providecommand \@ifxundefined [1]{%
 \ifx #1\undefined \expandafter \@firstoftwo
 \else \expandafter \@secondoftwo
\fi
}%
\providecommand \@ifnum [1]{%
 \ifnum #1\expandafter \@firstoftwo
 \else \expandafter \@secondoftwo
\fi
}%
\providecommand \enquote [1]{``#1''}%
\providecommand \bibnamefont  [1]{#1}%
\providecommand \bibfnamefont [1]{#1}%
\providecommand \citenamefont [1]{#1}%
\providecommand\href[0]{\@sanitize\@href}%
\providecommand\@href[1]{\endgroup\@@startlink{#1}\endgroup\@@href}%
\providecommand\@@href[1]{#1\@@endlink}%
\providecommand \@sanitize [0]{\begingroup\catcode`\&12\catcode`\#12\relax}%
\@ifxundefined \pdfoutput {\@firstoftwo}{%
 \@ifnum{\z@=\pdfoutput}{\@firstoftwo}{\@secondoftwo}%
}{%
 \providecommand\@@startlink[1]{\leavevmode\special{html:<a href="#1">}}%
 \providecommand\@@endlink[0]{\special{html:</a>}}%
}{%
 \providecommand\@@startlink[1]{%
  \leavevmode
  \pdfstartlink
   attr{/Border[0 0 1 ]/H/I/C[0 1 1]}%
   user{/Subtype/Link/A<</Type/Action/S/URI/URI(#1)>>}%
  \relax
 }%
 \providecommand\@@endlink[0]{\pdfendlink}%
}%
\providecommand \url  [0]{\begingroup\@sanitize \@url }%
\providecommand \@url [1]{\endgroup\@href {#1}{\urlprefix}}%
\providecommand \urlprefix [0]{URL }%
\providecommand \Eprint[0]{\href }%
\@ifxundefined \urlstyle {%
  \providecommand \doi [1]{doi:\discretionary{}{}{}#1}%
}{%
  \providecommand \doi [0]{doi:\discretionary{}{}{}\begingroup
  \urlstyle{rm}\Url }%
}%
\providecommand \doibase [0]{http://dx.doi.org/}%
\providecommand \Doi[1]{\href{\doibase#1}}%
\providecommand \bibAnnote [3]{%
  \BibitemShut{#1}%
  \begin{quotation}\noindent
    \textsc{Key:}\ #2\\\textsc{Annotation:}\ #3%
  \end{quotation}%
}%
\providecommand \bibAnnoteFile [2]{%
  \IfFileExists{#2}{\bibAnnote {#1} {#2} {\input{#2}}}{}%
}%
\providecommand \typeout [0]{\immediate \write \m@ne }%
\providecommand \selectlanguage [0]{\@gobble}%
\providecommand \bibinfo [0]{\@secondoftwo}%
\providecommand \bibfield [0]{\@secondoftwo}%
\providecommand \translation [1]{[#1]}%
\providecommand \BibitemOpen[0]{}%
\providecommand \bibitemStop [0]{}%
\providecommand \bibitemNoStop [0]{.\EOS\space}%
\providecommand \EOS [0]{\spacefactor3000\relax}%
\providecommand \BibitemShut [1]{\csname bibitem#1\endcsname}%
\bibitem{Bergstrom:1999jc}%
  \BibitemOpen
  \bibfield{author}{%
  \bibinfo {author} {\bibfnamefont{L.}~\bibnamefont{Bergstrom}}, \bibinfo
  {author} {\bibfnamefont{J.}~\bibnamefont{Edsjo}},\ and\ \bibinfo {author}
  {\bibfnamefont{P.}~\bibnamefont{Ullio}},\ }%
  \bibfield{journal}{%
  \Doi{10.1086/307975}{\bibinfo {journal} {Astrophys. J.}}\ }%
  \textbf{\bibinfo {volume} {526}},\ \bibinfo {pages} {215} (\bibinfo {year}
  {1999}),\
  \Eprint{http://arxiv.org/abs/astro-ph/9902012}{arXiv:astro-ph/9902012
  [astro-ph]}%
  \bibAnnoteFile{NoStop}{Bergstrom:1999jc}%
\bibitem{Donato:2003xg}%
  \BibitemOpen
  \bibfield{author}{%
  \bibinfo {author} {\bibfnamefont{F.}~\bibnamefont{Donato}}, \bibinfo {author}
  {\bibfnamefont{N.}~\bibnamefont{Fornengo}}, \bibinfo {author}
  {\bibfnamefont{D.}~\bibnamefont{Maurin}},\ and\ \bibinfo {author}
  {\bibfnamefont{P.}~\bibnamefont{Salati}},\ }%
  \bibfield{journal}{%
  \Doi{10.1103/PhysRevD.69.063501}{\bibinfo {journal} {Phys. Rev.}}\ }%
  \textbf{\bibinfo {volume} {D69}},\ \bibinfo {pages} {063501} (\bibinfo {year}
  {2004}),\
  \Eprint{http://arxiv.org/abs/astro-ph/0306207}{arXiv:astro-ph/0306207
  [astro-ph]}%
  \bibAnnoteFile{NoStop}{Donato:2003xg}%
\bibitem{Bringmann:2006im}%
  \BibitemOpen
  \bibfield{author}{%
  \bibinfo {author} {\bibfnamefont{T.}~\bibnamefont{Bringmann}}\ and\ \bibinfo
  {author} {\bibfnamefont{P.}~\bibnamefont{Salati}},\ }%
  \bibfield{journal}{%
  \Doi{10.1103/PhysRevD.75.083006}{\bibinfo {journal} {Phys. Rev.}}\ }%
  \textbf{\bibinfo {volume} {D75}},\ \bibinfo {pages} {083006} (\bibinfo {year}
  {2007}),\
  \Eprint{http://arxiv.org/abs/astro-ph/0612514}{arXiv:astro-ph/0612514
  [astro-ph]}%
  \bibAnnoteFile{NoStop}{Bringmann:2006im}%
\bibitem{Donato:2008jk}%
  \BibitemOpen
  \bibfield{author}{%
  \bibinfo {author} {\bibfnamefont{F.}~\bibnamefont{Donato}}, \bibinfo {author}
  {\bibfnamefont{D.}~\bibnamefont{Maurin}}, \bibinfo {author}
  {\bibfnamefont{P.}~\bibnamefont{Brun}}, \bibinfo {author}
  {\bibfnamefont{T.}~\bibnamefont{Delahaye}},\ and\ \bibinfo {author}
  {\bibfnamefont{P.}~\bibnamefont{Salati}},\ }%
  \bibfield{journal}{%
  \Doi{10.1103/PhysRevLett.102.071301}{\bibinfo {journal} {Phys. Rev. Lett.}}\
  }%
  \textbf{\bibinfo {volume} {102}},\ \bibinfo {pages} {071301} (\bibinfo {year}
  {2009}),\ \Eprint{http://arxiv.org/abs/0810.5292}{arXiv:0810.5292
  [astro-ph]}%
  \bibAnnoteFile{NoStop}{Donato:2008jk}%
\bibitem{Fornengo:2013xda}%
  \BibitemOpen
  \bibfield{author}{%
  \bibinfo {author} {\bibfnamefont{N.}~\bibnamefont{Fornengo}}, \bibinfo
  {author} {\bibfnamefont{L.}~\bibnamefont{Maccione}},\ and\ \bibinfo {author}
  {\bibfnamefont{A.}~\bibnamefont{Vittino}},\ }%
  \bibfield{journal}{%
  \Doi{10.1088/1475-7516/2014/04/003}{\bibinfo {journal} {JCAP}}\ }%
  \textbf{\bibinfo {volume} {1404}},\ \bibinfo {pages} {003} (\bibinfo {year}
  {2014}),\ \Eprint{http://arxiv.org/abs/1312.3579}{arXiv:1312.3579 [hep-ph]}%
  \bibAnnoteFile{NoStop}{Fornengo:2013xda}%
\bibitem{Hooper:2014ysa}%
  \BibitemOpen
  \bibfield{author}{%
  \bibinfo {author} {\bibfnamefont{D.}~\bibnamefont{Hooper}}, \bibinfo {author}
  {\bibfnamefont{T.}~\bibnamefont{Linden}},\ and\ \bibinfo {author}
  {\bibfnamefont{P.}~\bibnamefont{Mertsch}},\ }%
  \bibfield{journal}{%
  \Doi{10.1088/1475-7516/2015/03/021}{\bibinfo {journal} {JCAP}}\ }%
  \textbf{\bibinfo {volume} {1503}},\ \bibinfo {pages} {021} (\bibinfo {year}
  {2015}),\ \Eprint{http://arxiv.org/abs/1410.1527}{arXiv:1410.1527
  [astro-ph.HE]}%
  \bibAnnoteFile{NoStop}{Hooper:2014ysa}%
\bibitem{Pettorino:2014sua}%
  \BibitemOpen
  \bibfield{author}{%
  \bibinfo {author} {\bibfnamefont{V.}~\bibnamefont{Pettorino}}, \bibinfo
  {author} {\bibfnamefont{G.}~\bibnamefont{Busoni}}, \bibinfo {author}
  {\bibfnamefont{A.}~\bibnamefont{De~Simone}}, \bibinfo {author}
  {\bibfnamefont{E.}~\bibnamefont{Morgante}}, \bibinfo {author}
  {\bibfnamefont{A.}~\bibnamefont{Riotto}},\ and\ \bibinfo {author}
  {\bibfnamefont{W.}~\bibnamefont{Xue}},\ }%
  \bibfield{journal}{%
  \Doi{10.1088/1475-7516/2014/10/078}{\bibinfo {journal} {JCAP}}\ }%
  \textbf{\bibinfo {volume} {1410}},\ \bibinfo {pages} {078} (\bibinfo {year}
  {2014}),\ \Eprint{http://arxiv.org/abs/1406.5377}{arXiv:1406.5377 [hep-ph]}%
  \bibAnnoteFile{NoStop}{Pettorino:2014sua}%
\bibitem{Boudaud:2014qra}%
  \BibitemOpen
  \bibfield{author}{%
  \bibinfo {author} {\bibfnamefont{M.}~\bibnamefont{Boudaud}}, \bibinfo
  {author} {\bibfnamefont{M.}~\bibnamefont{Cirelli}}, \bibinfo {author}
  {\bibfnamefont{G.}~\bibnamefont{Giesen}},\ and\ \bibinfo {author}
  {\bibfnamefont{P.}~\bibnamefont{Salati}},\ }%
  \bibfield{journal}{%
  \Doi{10.1088/1475-7516/2015/05/013}{\bibinfo {journal} {JCAP}}\ }%
  \textbf{\bibinfo {volume} {1505}},\ \bibinfo {pages} {013} (\bibinfo {year}
  {2015}),\ \Eprint{http://arxiv.org/abs/1412.5696}{arXiv:1412.5696
  [astro-ph.HE]}%
  \bibAnnoteFile{NoStop}{Boudaud:2014qra}%
\bibitem{Cembranos:2014wza}%
  \BibitemOpen
  \bibfield{author}{%
  \bibinfo {author} {\bibfnamefont{J.~A.~R.}\ \bibnamefont{Cembranos}},
  \bibinfo {author} {\bibfnamefont{V.}~\bibnamefont{Gammaldi}},\ and\ \bibinfo
  {author} {\bibfnamefont{A.~L.}\ \bibnamefont{Maroto}},\ }%
  \bibfield{journal}{%
  \Doi{10.1088/1475-7516/2015/03/041}{\bibinfo {journal} {JCAP}}\ }%
  \textbf{\bibinfo {volume} {1503}},\ \bibinfo {pages} {041} (\bibinfo {year}
  {2015}),\ \Eprint{http://arxiv.org/abs/1410.6689}{arXiv:1410.6689
  [astro-ph.HE]}%
  \bibAnnoteFile{NoStop}{Cembranos:2014wza}%
\bibitem{Cirelli:2014lwa}%
  \BibitemOpen
  \bibfield{author}{%
  \bibinfo {author} {\bibfnamefont{M.}~\bibnamefont{Cirelli}}, \bibinfo
  {author} {\bibfnamefont{D.}~\bibnamefont{Gaggero}}, \bibinfo {author}
  {\bibfnamefont{G.}~\bibnamefont{Giesen}}, \bibinfo {author}
  {\bibfnamefont{M.}~\bibnamefont{Taoso}},\ and\ \bibinfo {author}
  {\bibfnamefont{A.}~\bibnamefont{Urbano}},\ }%
  \bibfield{journal}{%
  \Doi{10.1088/1475-7516/2014/12/045}{\bibinfo {journal} {JCAP}}\ }%
  \textbf{\bibinfo {volume} {1412}},\ \bibinfo {pages} {045} (\bibinfo {year}
  {2014}),\ \Eprint{http://arxiv.org/abs/1407.2173}{arXiv:1407.2173 [hep-ph]}%
  \bibAnnoteFile{NoStop}{Cirelli:2014lwa}%
\bibitem{Bringmann:2014lpa}%
  \BibitemOpen
  \bibfield{author}{%
  \bibinfo {author} {\bibfnamefont{T.}~\bibnamefont{Bringmann}}, \bibinfo
  {author} {\bibfnamefont{M.}~\bibnamefont{Vollmann}},\ and\ \bibinfo {author}
  {\bibfnamefont{C.}~\bibnamefont{Weniger}},\ }%
  \bibfield{journal}{%
  \Doi{10.1103/PhysRevD.90.123001}{\bibinfo {journal} {Phys. Rev.}}\ }%
  \textbf{\bibinfo {volume} {D90}},\ \bibinfo {pages} {123001} (\bibinfo {year}
  {2014}),\ \Eprint{http://arxiv.org/abs/1406.6027}{arXiv:1406.6027
  [astro-ph.HE]}%
  \bibAnnoteFile{NoStop}{Bringmann:2014lpa}%
\bibitem{Giesen:2015ufa}%
  \BibitemOpen
  \bibfield{author}{%
  \bibinfo {author} {\bibfnamefont{G.}~\bibnamefont{Giesen}}, \bibinfo {author}
  {\bibfnamefont{M.}~\bibnamefont{Boudaud}}, \bibinfo {author}
  {\bibfnamefont{Y.}~\bibnamefont{Genolini}}, \bibinfo {author}
  {\bibfnamefont{V.}~\bibnamefont{Poulin}}, \bibinfo {author}
  {\bibfnamefont{M.}~\bibnamefont{Cirelli}}, \bibinfo {author}
  {\bibfnamefont{P.}~\bibnamefont{Salati}},\ and\ \bibinfo {author}
  {\bibfnamefont{P.~D.}\ \bibnamefont{Serpico}},\ }%
  \bibfield{journal}{%
  \Doi{10.1088/1475-7516/2015/09/023, 10.1088/1475-7516/2015/9/023}{\bibinfo
  {journal} {JCAP}}\ }%
  \textbf{\bibinfo {volume} {1509}},\ \bibinfo {pages} {023} (\bibinfo {year}
  {2015}),\ \Eprint{http://arxiv.org/abs/1504.04276}{arXiv:1504.04276
  [astro-ph.HE]}%
  \bibAnnoteFile{NoStop}{Giesen:2015ufa}%
\bibitem{Evoli:2015vaa}%
  \BibitemOpen
  \bibfield{author}{%
  \bibinfo {author} {\bibfnamefont{C.}~\bibnamefont{Evoli}}, \bibinfo {author}
  {\bibfnamefont{D.}~\bibnamefont{Gaggero}},\ and\ \bibinfo {author}
  {\bibfnamefont{D.}~\bibnamefont{Grasso}},\ }%
  \bibfield{journal}{%
  \Doi{10.1088/1475-7516/2015/12/039}{\bibinfo {journal} {JCAP}}\ }%
  \textbf{\bibinfo {volume} {1512}},\ \bibinfo {pages} {039} (\bibinfo {year}
  {2015}),\ \Eprint{http://arxiv.org/abs/1504.05175}{arXiv:1504.05175
  [astro-ph.HE]}%
  \bibAnnoteFile{NoStop}{Evoli:2015vaa}%
\bibitem{Aguilar:2016kjl}%
  \BibitemOpen
  \bibfield{author}{%
  \bibinfo {author} {\bibfnamefont{M.}~\bibnamefont{Aguilar}} \emph{et~al.}
  (\bibinfo {collaboration} {AMS}),\ }%
  \bibfield{journal}{%
  \bibinfo {journal} {Phys. Rev. Lett.}\ }%
  \textbf{\bibinfo {volume} {117}} (\bibinfo {year} {2016}),\ \doi{\bibinfo
  {doi} {10.1103/PhysRevLett.117.091103}}%
  \bibAnnoteFile{NoStop}{Aguilar:2016kjl}%
\bibitem{Aguilar:2016vqr}%
  \BibitemOpen
  \bibfield{author}{%
  \bibinfo {author} {\bibfnamefont{M.}~\bibnamefont{Aguilar}} \emph{et~al.}
  (\bibinfo {collaboration} {AMS}),\ }%
  \bibfield{journal}{%
  \Doi{10.1103/PhysRevLett.117.231102}{\bibinfo {journal} {Phys. Rev. Lett.}}\
  }%
  \textbf{\bibinfo {volume} {117}},\ \bibinfo {pages} {231102} (\bibinfo {year}
  {2016})%
  \bibAnnoteFile{NoStop}{Aguilar:2016vqr}%
\bibitem{Johannesson_CR_Propagation_2016}%
  \BibitemOpen
  \bibfield{author}{%
  \bibinfo {author} {\bibfnamefont{G.}~\bibnamefont{Johannesson}}
  \emph{et~al.},\ }%
  \bibfield{journal}{%
  \Doi{10.3847/0004-637X/824/1/16}{\bibinfo {journal} {Astrophys. J.}}\ }%
  \textbf{\bibinfo {volume} {824}},\ \bibinfo {pages} {16} (\bibinfo {year}
  {2016}),\ \Eprint{http://arxiv.org/abs/1602.02243}{arXiv:1602.02243
  [astro-ph.HE]}%
  \bibAnnoteFile{NoStop}{Johannesson_CR_Propagation_2016}%
\bibitem{Yuan:2017ozr}%
  \BibitemOpen
  \bibfield{author}{%
  \bibinfo {author} {\bibfnamefont{Q.}~\bibnamefont{Yuan}}, \bibinfo {author}
  {\bibfnamefont{S.-J.}\ \bibnamefont{Lin}}, \bibinfo {author}
  {\bibfnamefont{K.}~\bibnamefont{Fang}},\ and\ \bibinfo {author}
  {\bibfnamefont{X.-J.}\ \bibnamefont{Bi}}}%
   (\bibinfo {year} {2017}),\
  \Eprint{http://arxiv.org/abs/1701.06149}{arXiv:1701.06149 [astro-ph.HE]}%
  \bibAnnoteFile{NoStop}{Yuan:2017ozr}%
\bibitem{Jin:2017iwg}%
  \BibitemOpen
  \bibfield{author}{%
  \bibinfo {author} {\bibfnamefont{H.-B.}\ \bibnamefont{Jin}}, \bibinfo
  {author} {\bibfnamefont{Y.-L.}\ \bibnamefont{Wu}},\ and\ \bibinfo {author}
  {\bibfnamefont{Y.-F.}\ \bibnamefont{Zhou}}}%
   (\bibinfo {year} {2017}),\
  \Eprint{http://arxiv.org/abs/1701.02213}{arXiv:1701.02213 [hep-ph]}%
  \bibAnnoteFile{NoStop}{Jin:2017iwg}%
\bibitem{Lin:2016ezz}%
  \BibitemOpen
  \bibfield{author}{%
  \bibinfo {author} {\bibfnamefont{S.-J.}\ \bibnamefont{Lin}}, \bibinfo
  {author} {\bibfnamefont{X.-J.}\ \bibnamefont{Bi}}, \bibinfo {author}
  {\bibfnamefont{J.}~\bibnamefont{Feng}}, \bibinfo {author}
  {\bibfnamefont{P.-F.}\ \bibnamefont{Yin}},\ and\ \bibinfo {author}
  {\bibfnamefont{Z.-H.}\ \bibnamefont{Yu}}}%
   (\bibinfo {year} {2016}),\
  \Eprint{http://arxiv.org/abs/1612.04001}{arXiv:1612.04001 [astro-ph.HE]}%
  \bibAnnoteFile{NoStop}{Lin:2016ezz}%
\bibitem{Korsmeier:2016kha}%
  \BibitemOpen
  \bibfield{author}{%
  \bibinfo {author} {\bibfnamefont{M.}~\bibnamefont{Korsmeier}}\ and\ \bibinfo
  {author} {\bibfnamefont{A.}~\bibnamefont{Cuoco}},\ }%
  \bibfield{journal}{%
  \Doi{10.1103/PhysRevD.94.123019}{\bibinfo {journal} {Phys. Rev.}}\ }%
  \textbf{\bibinfo {volume} {D94}},\ \bibinfo {pages} {123019} (\bibinfo {year}
  {2016}),\ \Eprint{http://arxiv.org/abs/1607.06093}{arXiv:1607.06093
  [astro-ph.HE]}%
  \bibAnnoteFile{NoStop}{Korsmeier:2016kha}%
\bibitem{Navarro:1995iw}%
  \BibitemOpen
  \bibfield{author}{%
  \bibinfo {author} {\bibfnamefont{J.~F.}\ \bibnamefont{Navarro}}, \bibinfo
  {author} {\bibfnamefont{C.~S.}\ \bibnamefont{Frenk}},\ and\ \bibinfo {author}
  {\bibfnamefont{S.~D.~M.}\ \bibnamefont{White}},\ }%
  \bibfield{journal}{%
  \Doi{10.1086/177173}{\bibinfo {journal} {Astrophys. J.}}\ }%
  \textbf{\bibinfo {volume} {462}},\ \bibinfo {pages} {563} (\bibinfo {year}
  {1996}),\
  \Eprint{http://arxiv.org/abs/astro-ph/9508025}{arXiv:astro-ph/9508025
  [astro-ph]}%
  \bibAnnoteFile{NoStop}{Navarro:1995iw}%
\bibitem{Salucci:2010qr}%
  \BibitemOpen
  \bibfield{author}{%
  \bibinfo {author} {\bibfnamefont{P.}~\bibnamefont{Salucci}}, \bibinfo
  {author} {\bibfnamefont{F.}~\bibnamefont{Nesti}}, \bibinfo {author}
  {\bibfnamefont{G.}~\bibnamefont{Gentile}},\ and\ \bibinfo {author}
  {\bibfnamefont{C.~F.}\ \bibnamefont{Martins}},\ }%
  \bibfield{journal}{%
  \Doi{10.1051/0004-6361/201014385}{\bibinfo {journal} {Astron. Astrophys.}}\
  }%
  \textbf{\bibinfo {volume} {523}},\ \bibinfo {pages} {A83} (\bibinfo {year}
  {2010}),\ \Eprint{http://arxiv.org/abs/1003.3101}{arXiv:1003.3101
  [astro-ph.GA]}%
  \bibAnnoteFile{NoStop}{Salucci:2010qr}%
\bibitem{Burkert:1995yz}%
  \BibitemOpen
  \bibfield{author}{%
  \bibinfo {author} {\bibfnamefont{A.}~\bibnamefont{Burkert}},\ }%
  \bibfield{journal}{%
  \Doi{10.1086/309560}{\bibinfo {journal} {IAU Symp.}}\ }%
  \textbf{\bibinfo {volume} {171}},\ \bibinfo {pages} {175} (\bibinfo {year}
  {1996}),\ \bibinfo {note} {[Astrophys. J.447,L25(1995)]},\
  \Eprint{http://arxiv.org/abs/astro-ph/9504041}{arXiv:astro-ph/9504041
  [astro-ph]}%
  \bibAnnoteFile{NoStop}{Burkert:1995yz}%
\bibitem{Cirelli:2010xx}%
  \BibitemOpen
  \bibfield{author}{%
  \bibinfo {author} {\bibfnamefont{M.}~\bibnamefont{Cirelli}}, \bibinfo
  {author} {\bibfnamefont{G.}~\bibnamefont{Corcella}}, \bibinfo {author}
  {\bibfnamefont{A.}~\bibnamefont{Hektor}}, \bibinfo {author}
  {\bibfnamefont{G.}~\bibnamefont{Hutsi}}, \bibinfo {author}
  {\bibfnamefont{M.}~\bibnamefont{Kadastik}}, \bibinfo {author}
  {\bibfnamefont{P.}~\bibnamefont{Panci}}, \bibinfo {author}
  {\bibfnamefont{M.}~\bibnamefont{Raidal}}, \bibinfo {author}
  {\bibfnamefont{F.}~\bibnamefont{Sala}},\ and\ \bibinfo {author}
  {\bibfnamefont{A.}~\bibnamefont{Strumia}},\ }%
  \bibfield{journal}{%
  \Doi{10.1088/1475-7516/2012/10/E01, 10.1088/1475-7516/2011/03/051}{\bibinfo
  {journal} {JCAP}}\ }%
  \textbf{\bibinfo {volume} {1103}},\ \bibinfo {pages} {051} (\bibinfo {year}
  {2011}),\ \bibinfo {note} {[Erratum: JCAP1210,E01(2012)]},\
  \Eprint{http://arxiv.org/abs/1012.4515}{arXiv:1012.4515 [hep-ph]}%
  \bibAnnoteFile{NoStop}{Cirelli:2010xx}%
\bibitem{StrongMoskalenko_CR_rewview_2007}%
  \BibitemOpen
  \bibfield{author}{%
  \bibinfo {author} {\bibfnamefont{A.~W.}\ \bibnamefont{Strong}}, \bibinfo
  {author} {\bibfnamefont{I.~V.}\ \bibnamefont{Moskalenko}},\ and\ \bibinfo
  {author} {\bibfnamefont{V.~S.}\ \bibnamefont{Ptuskin}},\ }%
  \bibfield{journal}{%
  \Doi{10.1146/annurev.nucl.57.090506.123011}{\bibinfo {journal} {Ann. Rev.
  Nucl. Part. Sci.}}\ }%
  \textbf{\bibinfo {volume} {57}},\ \bibinfo {pages} {285} (\bibinfo {year}
  {2007}),\
  \Eprint{http://arxiv.org/abs/astro-ph/0701517}{arXiv:astro-ph/0701517
  [astro-ph]}%
  \bibAnnoteFile{NoStop}{StrongMoskalenko_CR_rewview_2007}%
\bibitem{Strong:1998fr}%
  \BibitemOpen
  \bibfield{author}{%
  \bibinfo {author} {\bibfnamefont{A.~W.}\ \bibnamefont{Strong}}, \bibinfo
  {author} {\bibfnamefont{I.~V.}\ \bibnamefont{Moskalenko}},\ and\ \bibinfo
  {author} {\bibfnamefont{O.}~\bibnamefont{Reimer}},\ }%
  \bibfield{journal}{%
  \Doi{10.1086/309038}{\bibinfo {journal} {Astrophys. J.}}\ }%
  \textbf{\bibinfo {volume} {537}},\ \bibinfo {pages} {763} (\bibinfo {year}
  {2000}),\ \bibinfo {note} {[Erratum: Astrophys. J.541,1109(2000)]},\
  \Eprint{http://arxiv.org/abs/astro-ph/9811296}{arXiv:astro-ph/9811296
  [astro-ph]}%
  \bibAnnoteFile{NoStop}{Strong:1998fr}%
\bibitem{Strong:2015zva}%
  \BibitemOpen
  \bibfield{author}{%
  \bibinfo {author} {\bibfnamefont{A.~W.}\ \bibnamefont{Strong}}}%
   (\bibinfo {year} {2015}),\
  \Eprint{http://arxiv.org/abs/1507.05020}{arXiv:1507.05020 [astro-ph.HE]}%
  \bibAnnoteFile{NoStop}{Strong:2015zva}%
\bibitem{Moskalenko:2001ya}%
  \BibitemOpen
  \bibfield{author}{%
  \bibinfo {author} {\bibfnamefont{I.~V.}\ \bibnamefont{Moskalenko}}, \bibinfo
  {author} {\bibfnamefont{A.~W.}\ \bibnamefont{Strong}}, \bibinfo {author}
  {\bibfnamefont{J.~F.}\ \bibnamefont{Ormes}},\ and\ \bibinfo {author}
  {\bibfnamefont{M.~S.}\ \bibnamefont{Potgieter}},\ }%
  \bibfield{journal}{%
  \Doi{10.1086/324402}{\bibinfo {journal} {Astrophys. J.}}\ }%
  \textbf{\bibinfo {volume} {565}},\ \bibinfo {pages} {280} (\bibinfo {year}
  {2002}),\
  \Eprint{http://arxiv.org/abs/astro-ph/0106567}{arXiv:astro-ph/0106567
  [astro-ph]}%
  \bibAnnoteFile{NoStop}{Moskalenko:2001ya}%
\bibitem{Aguilar_AMS_Proton_2015}%
  \BibitemOpen
  \bibfield{author}{%
  \bibinfo {author} {\bibfnamefont{M.}~\bibnamefont{Aguilar}} \emph{et~al.}
  (\bibinfo {collaboration} {AMS}),\ }%
  \bibfield{journal}{%
  \Doi{10.1103/PhysRevLett.114.171103}{\bibinfo {journal} {Phys. Rev. Lett.}}\
  }%
  \textbf{\bibinfo {volume} {114}},\ \bibinfo {pages} {171103} (\bibinfo {year}
  {2015})%
  \bibAnnoteFile{NoStop}{Aguilar_AMS_Proton_2015}%
\bibitem{Aguilar_AMS_Helium_2015}%
  \BibitemOpen
  \bibfield{author}{%
  \bibinfo {author} {\bibfnamefont{M.}~\bibnamefont{Aguilar}} \emph{et~al.}
  (\bibinfo {collaboration} {AMS}),\ }%
  \bibfield{journal}{%
  \Doi{10.1103/PhysRevLett.115.211101}{\bibinfo {journal} {Phys. Rev. Lett.}}\
  }%
  \textbf{\bibinfo {volume} {115}},\ \bibinfo {pages} {211101} (\bibinfo {year}
  {2015})%
  \bibAnnoteFile{NoStop}{Aguilar_AMS_Helium_2015}%
\bibitem{Yoon_CREAM_CR_ProtonHelium_2011}%
  \BibitemOpen
  \bibfield{author}{%
  \bibinfo {author} {\bibfnamefont{Y.~S.}\ \bibnamefont{Yoon}} \emph{et~al.},\
  }%
  \bibfield{journal}{%
  \Doi{10.1088/0004-637X/728/2/122}{\bibinfo {journal} {Astrophys. J.}}\ }%
  \textbf{\bibinfo {volume} {728}},\ \bibinfo {pages} {122} (\bibinfo {year}
  {2011}),\ \Eprint{http://arxiv.org/abs/1102.2575}{arXiv:1102.2575
  [astro-ph.HE]}%
  \bibAnnoteFile{NoStop}{Yoon_CREAM_CR_ProtonHelium_2011}%
\bibitem{Stone_VOYAGER_CR_LIS_FLUX_2013}%
  \BibitemOpen
  \bibfield{author}{%
  \bibinfo {author} {\bibfnamefont{E.~C.}\ \bibnamefont{Stone}} \emph{et~al.},\
  }%
  \bibfield{journal}{%
  \bibinfo {journal} {Science}\ }%
  \textbf{\bibinfo {volume} {341}} (\bibinfo {year} {2013}),\ \doi{\bibinfo
  {doi} {10.1126/science.1236408}}%
  \bibAnnoteFile{NoStop}{Stone_VOYAGER_CR_LIS_FLUX_2013}%
\bibitem{Feroz_MultiNest_2008}%
  \BibitemOpen
  \bibfield{author}{%
  \bibinfo {author} {\bibfnamefont{F.}~\bibnamefont{Feroz}}, \bibinfo {author}
  {\bibfnamefont{M.~P.}\ \bibnamefont{Hobson}},\ and\ \bibinfo {author}
  {\bibfnamefont{M.}~\bibnamefont{Bridges}},\ }%
  \bibfield{journal}{%
  \Doi{10.1111/j.1365-2966.2009.14548.x}{\bibinfo {journal} {Mon. Not. Roy.
  Astron. Soc.}}\ }%
  \textbf{\bibinfo {volume} {398}},\ \bibinfo {pages} {1601} (\bibinfo {year}
  {2009}),\ \Eprint{http://arxiv.org/abs/0809.3437}{arXiv:0809.3437
  [astro-ph]}%
  \bibAnnoteFile{NoStop}{Feroz_MultiNest_2008}%
\bibitem{Gordon:2013vta}%
  \BibitemOpen
  \bibfield{author}{%
  \bibinfo {author} {\bibfnamefont{C.}~\bibnamefont{Gordon}}\ and\ \bibinfo
  {author} {\bibfnamefont{O.}~\bibnamefont{Macias}},\ }%
  \bibfield{journal}{%
  \Doi{10.1103/PhysRevD.88.083521, 10.1103/PhysRevD.89.049901}{\bibinfo
  {journal} {Phys. Rev.}}\ }%
  \textbf{\bibinfo {volume} {D88}},\ \bibinfo {pages} {083521} (\bibinfo {year}
  {2013}),\ \bibinfo {note} {[Erratum: Phys. Rev.D89,no.4,049901(2014)]},\
  \Eprint{http://arxiv.org/abs/1306.5725}{arXiv:1306.5725 [astro-ph.HE]}%
  \bibAnnoteFile{NoStop}{Gordon:2013vta}%
\bibitem{Abazajian:2014fta}%
  \BibitemOpen
  \bibfield{author}{%
  \bibinfo {author} {\bibfnamefont{K.~N.}\ \bibnamefont{Abazajian}}, \bibinfo
  {author} {\bibfnamefont{N.}~\bibnamefont{Canac}}, \bibinfo {author}
  {\bibfnamefont{S.}~\bibnamefont{Horiuchi}},\ and\ \bibinfo {author}
  {\bibfnamefont{M.}~\bibnamefont{Kaplinghat}},\ }%
  \bibfield{journal}{%
  \Doi{10.1103/PhysRevD.90.023526}{\bibinfo {journal} {Phys. Rev.}}\ }%
  \textbf{\bibinfo {volume} {D90}},\ \bibinfo {pages} {023526} (\bibinfo {year}
  {2014}),\ \Eprint{http://arxiv.org/abs/1402.4090}{arXiv:1402.4090
  [astro-ph.HE]}%
  \bibAnnoteFile{NoStop}{Abazajian:2014fta}%
\bibitem{Daylan:2014rsa}%
  \BibitemOpen
  \bibfield{author}{%
  \bibinfo {author} {\bibfnamefont{T.}~\bibnamefont{Daylan}}, \bibinfo {author}
  {\bibfnamefont{D.~P.}\ \bibnamefont{Finkbeiner}}, \bibinfo {author}
  {\bibfnamefont{D.}~\bibnamefont{Hooper}}, \bibinfo {author}
  {\bibfnamefont{T.}~\bibnamefont{Linden}}, \bibinfo {author}
  {\bibfnamefont{S.~K.~N.}\ \bibnamefont{Portillo}}, \bibinfo {author}
  {\bibfnamefont{N.~L.}\ \bibnamefont{Rodd}},\ and\ \bibinfo {author}
  {\bibfnamefont{T.~R.}\ \bibnamefont{Slatyer}},\ }%
  \bibfield{journal}{%
  \Doi{10.1016/j.dark.2015.12.005}{\bibinfo {journal} {Phys. Dark Univ.}}\ }%
  \textbf{\bibinfo {volume} {12}},\ \bibinfo {pages} {1} (\bibinfo {year}
  {2016}),\ \Eprint{http://arxiv.org/abs/1402.6703}{arXiv:1402.6703
  [astro-ph.HE]}%
  \bibAnnoteFile{NoStop}{Daylan:2014rsa}%
\bibitem{TheFermi-LAT:2015kwa}%
  \BibitemOpen
  \bibfield{author}{%
  \bibinfo {author} {\bibfnamefont{M.}~\bibnamefont{Ajello}} \emph{et~al.}
  (\bibinfo {collaboration} {Fermi-LAT}),\ }%
  \bibfield{journal}{%
  \Doi{10.3847/0004-637X/819/1/44}{\bibinfo {journal} {Astrophys. J.}}\ }%
  \textbf{\bibinfo {volume} {819}},\ \bibinfo {pages} {44} (\bibinfo {year}
  {2016}),\ \Eprint{http://arxiv.org/abs/1511.02938}{arXiv:1511.02938
  [astro-ph.HE]}%
  \bibAnnoteFile{NoStop}{TheFermi-LAT:2015kwa}%
\bibitem{Calore:2014nla}%
  \BibitemOpen
  \bibfield{author}{%
  \bibinfo {author} {\bibfnamefont{F.}~\bibnamefont{Calore}}, \bibinfo {author}
  {\bibfnamefont{I.}~\bibnamefont{Cholis}}, \bibinfo {author}
  {\bibfnamefont{C.}~\bibnamefont{McCabe}},\ and\ \bibinfo {author}
  {\bibfnamefont{C.}~\bibnamefont{Weniger}},\ }%
  \bibfield{journal}{%
  \Doi{10.1103/PhysRevD.91.063003}{\bibinfo {journal} {Phys. Rev.}}\ }%
  \textbf{\bibinfo {volume} {D91}},\ \bibinfo {pages} {063003} (\bibinfo {year}
  {2015}),\ \Eprint{http://arxiv.org/abs/1411.4647}{arXiv:1411.4647 [hep-ph]}%
  \bibAnnoteFile{NoStop}{Calore:2014nla}%
\bibitem{Adriani:2010rc}%
  \BibitemOpen
  \bibfield{author}{%
  \bibinfo {author} {\bibfnamefont{O.}~\bibnamefont{Adriani}} \emph{et~al.}
  (\bibinfo {collaboration} {PAMELA}),\ }%
  \bibfield{journal}{%
  \Doi{10.1103/PhysRevLett.105.121101}{\bibinfo {journal} {Phys. Rev. Lett.}}\
  }%
  \textbf{\bibinfo {volume} {105}},\ \bibinfo {pages} {121101} (\bibinfo {year}
  {2010}),\ \Eprint{http://arxiv.org/abs/1007.0821}{arXiv:1007.0821
  [astro-ph.HE]}%
  \bibAnnoteFile{NoStop}{Adriani:2010rc}%
\bibitem{TanNg_AntiprotonParametrization_1983}%
  \BibitemOpen
  \bibfield{author}{%
  \bibinfo {author} {\bibfnamefont{L.~C.}\ \bibnamefont{Tan}}\ and\ \bibinfo
  {author} {\bibfnamefont{L.~K.}\ \bibnamefont{Ng}},\ }%
  \bibfield{journal}{%
  \Doi{10.1088/0305-4616/9/2/015}{\bibinfo {journal} {J. Phys.}}\ }%
  \textbf{\bibinfo {volume} {G9}},\ \bibinfo {pages} {227} (\bibinfo {year}
  {1983})%
  \bibAnnoteFile{NoStop}{TanNg_AntiprotonParametrization_1983}%
\bibitem{Mauro_Antiproton_Cross_Section_2014}%
  \BibitemOpen
  \bibfield{author}{%
  \bibinfo {author} {\bibfnamefont{M.}~\bibnamefont{di~Mauro}}, \bibinfo
  {author} {\bibfnamefont{F.}~\bibnamefont{Donato}}, \bibinfo {author}
  {\bibfnamefont{A.}~\bibnamefont{Goudelis}},\ and\ \bibinfo {author}
  {\bibfnamefont{P.~D.}\ \bibnamefont{Serpico}},\ }%
  \bibfield{journal}{%
  \Doi{10.1103/PhysRevD.90.085017}{\bibinfo {journal} {Phys. Rev.}}\ }%
  \textbf{\bibinfo {volume} {D90}},\ \bibinfo {pages} {085017} (\bibinfo {year}
  {2014}),\ \Eprint{http://arxiv.org/abs/1408.0288}{arXiv:1408.0288 [hep-ph]}%
  \bibAnnoteFile{NoStop}{Mauro_Antiproton_Cross_Section_2014}%
\bibitem{Kachelriess:2015wpa}%
  \BibitemOpen
  \bibfield{author}{%
  \bibinfo {author} {\bibfnamefont{M.}~\bibnamefont{Kachelriess}}, \bibinfo
  {author} {\bibfnamefont{I.~V.}\ \bibnamefont{Moskalenko}},\ and\ \bibinfo
  {author} {\bibfnamefont{S.~S.}\ \bibnamefont{Ostapchenko}},\ }%
  \bibfield{journal}{%
  \Doi{10.1088/0004-637X/803/2/54}{\bibinfo {journal} {Astrophys. J.}}\ }%
  \textbf{\bibinfo {volume} {803}},\ \bibinfo {pages} {54} (\bibinfo {year}
  {2015}),\ \Eprint{http://arxiv.org/abs/1502.04158}{arXiv:1502.04158
  [astro-ph.HE]}%
  \bibAnnoteFile{NoStop}{Kachelriess:2015wpa}%
\bibitem{Donato:2001ms}%
  \BibitemOpen
  \bibfield{author}{%
  \bibinfo {author} {\bibfnamefont{F.}~\bibnamefont{Donato}}, \bibinfo {author}
  {\bibfnamefont{D.}~\bibnamefont{Maurin}}, \bibinfo {author}
  {\bibfnamefont{P.}~\bibnamefont{Salati}}, \bibinfo {author}
  {\bibfnamefont{A.}~\bibnamefont{Barrau}}, \bibinfo {author}
  {\bibfnamefont{G.}~\bibnamefont{Boudoul}},\ and\ \bibinfo {author}
  {\bibfnamefont{R.}~\bibnamefont{Taillet}},\ }%
  \bibfield{journal}{%
  \Doi{10.1086/323684}{\bibinfo {journal} {Astrophys. J.}}\ }%
  \textbf{\bibinfo {volume} {563}},\ \bibinfo {pages} {172} (\bibinfo {year}
  {2001}),\
  \Eprint{http://arxiv.org/abs/astro-ph/0103150}{arXiv:astro-ph/0103150
  [astro-ph]}%
  \bibAnnoteFile{NoStop}{Donato:2001ms}%
\bibitem{Evoli:2011id}%
  \BibitemOpen
  \bibfield{author}{%
  \bibinfo {author} {\bibfnamefont{C.}~\bibnamefont{Evoli}}, \bibinfo {author}
  {\bibfnamefont{I.}~\bibnamefont{Cholis}}, \bibinfo {author}
  {\bibfnamefont{D.}~\bibnamefont{Grasso}}, \bibinfo {author}
  {\bibfnamefont{L.}~\bibnamefont{Maccione}},\ and\ \bibinfo {author}
  {\bibfnamefont{P.}~\bibnamefont{Ullio}},\ }%
  \bibfield{journal}{%
  \Doi{10.1103/PhysRevD.85.123511}{\bibinfo {journal} {Phys. Rev.}}\ }%
  \textbf{\bibinfo {volume} {D85}},\ \bibinfo {pages} {123511} (\bibinfo {year}
  {2012}),\ \Eprint{http://arxiv.org/abs/1108.0664}{arXiv:1108.0664
  [astro-ph.HE]}%
  \bibAnnoteFile{NoStop}{Evoli:2011id}%
\bibitem{Kappl:2014hha}%
  \BibitemOpen
  \bibfield{author}{%
  \bibinfo {author} {\bibfnamefont{R.}~\bibnamefont{Kappl}}\ and\ \bibinfo
  {author} {\bibfnamefont{M.~W.}\ \bibnamefont{Winkler}},\ }%
  \bibfield{journal}{%
  \Doi{10.1088/1475-7516/2014/09/051}{\bibinfo {journal} {JCAP}}\ }%
  \textbf{\bibinfo {volume} {1409}},\ \bibinfo {pages} {051} (\bibinfo {year}
  {2014}),\ \Eprint{http://arxiv.org/abs/1408.0299}{arXiv:1408.0299 [hep-ph]}%
  \bibAnnoteFile{NoStop}{Kappl:2014hha}%
\bibitem{Winkler:2017xor}%
  \BibitemOpen
  \bibfield{author}{%
  \bibinfo {author} {\bibfnamefont{M.~W.}\ \bibnamefont{Winkler}}}%
   (\bibinfo {year} {2017}),\
  \Eprint{http://arxiv.org/abs/1701.04866}{arXiv:1701.04866 [hep-ph]}%
  \bibAnnoteFile{NoStop}{Winkler:2017xor}%
\bibitem{Cholis_Solar_Modulation_2016}%
  \BibitemOpen
  \bibfield{author}{%
  \bibinfo {author} {\bibfnamefont{I.}~\bibnamefont{Cholis}}, \bibinfo {author}
  {\bibfnamefont{D.}~\bibnamefont{Hooper}},\ and\ \bibinfo {author}
  {\bibfnamefont{T.}~\bibnamefont{Linden}},\ }%
  \bibfield{journal}{%
  \Doi{10.1103/PhysRevD.93.043016}{\bibinfo {journal} {Phys. Rev.}}\ }%
  \textbf{\bibinfo {volume} {D93}},\ \bibinfo {pages} {043016} (\bibinfo {year}
  {2016}),\ \Eprint{http://arxiv.org/abs/1511.01507}{arXiv:1511.01507
  [astro-ph.SR]}%
  \bibAnnoteFile{NoStop}{Cholis_Solar_Modulation_2016}%
\bibitem{Corti:2015bqi}%
  \BibitemOpen
  \bibfield{author}{%
  \bibinfo {author} {\bibfnamefont{C.}~\bibnamefont{Corti}}, \bibinfo {author}
  {\bibfnamefont{V.}~\bibnamefont{Bindi}}, \bibinfo {author}
  {\bibfnamefont{C.}~\bibnamefont{Consolandi}},\ and\ \bibinfo {author}
  {\bibfnamefont{K.}~\bibnamefont{Whitman}},\ }%
  \bibfield{journal}{%
  \Doi{10.3847/0004-637X/829/1/8}{\bibinfo {journal} {Astrophys. J.}}\ }%
  \textbf{\bibinfo {volume} {829}},\ \bibinfo {pages} {8} (\bibinfo {year}
  {2016}),\ \Eprint{http://arxiv.org/abs/1511.08790}{arXiv:1511.08790
  [astro-ph.HE]}%
  \bibAnnoteFile{NoStop}{Corti:2015bqi}%
\bibitem{Ackermann:2015zua}%
  \BibitemOpen
  \bibfield{author}{%
  \bibinfo {author} {\bibfnamefont{M.}~\bibnamefont{Ackermann}} \emph{et~al.}
  (\bibinfo {collaboration} {Fermi-LAT}),\ }%
  \bibfield{journal}{%
  \Doi{10.1103/PhysRevLett.115.231301}{\bibinfo {journal} {Phys. Rev. Lett.}}\
  }%
  \textbf{\bibinfo {volume} {115}},\ \bibinfo {pages} {231301} (\bibinfo {year}
  {2015}),\ \Eprint{http://arxiv.org/abs/1503.02641}{arXiv:1503.02641
  [astro-ph.HE]}%
  \bibAnnoteFile{NoStop}{Ackermann:2015zua}%
\bibitem{Bonnivard:2015xpq}%
  \BibitemOpen
  \bibfield{author}{%
  \bibinfo {author} {\bibfnamefont{V.}~\bibnamefont{Bonnivard}} \emph{et~al.},\
  }%
  \bibfield{journal}{%
  \Doi{10.1093/mnras/stv1601}{\bibinfo {journal} {Mon. Not. Roy. Astron.
  Soc.}}\ }%
  \textbf{\bibinfo {volume} {453}},\ \bibinfo {pages} {849} (\bibinfo {year}
  {2015}),\ \Eprint{http://arxiv.org/abs/1504.02048}{arXiv:1504.02048
  [astro-ph.HE]}%
  \bibAnnoteFile{NoStop}{Bonnivard:2015xpq}%
\bibitem{Fermi-LAT:2016uux}%
  \BibitemOpen
  \bibfield{author}{%
  \bibinfo {author} {\bibfnamefont{A.}~\bibnamefont{Albert}} \emph{et~al.}
  (\bibinfo {collaboration} {DES, Fermi-LAT}),\ }%
  \bibfield{journal}{%
  \Doi{10.3847/1538-4357/834/2/110}{\bibinfo {journal} {Astrophys. J.}}\ }%
  \textbf{\bibinfo {volume} {834}},\ \bibinfo {pages} {110} (\bibinfo {year}
  {2017}),\ \Eprint{http://arxiv.org/abs/1611.03184}{arXiv:1611.03184
  [astro-ph.HE]}%
  \bibAnnoteFile{NoStop}{Fermi-LAT:2016uux}%
\bibitem{Cui_2016}%
  \BibitemOpen
  \bibfield{author}{%
  \bibinfo {author} {\bibfnamefont{M.-Y.}\ \bibnamefont{Cui}}, \bibinfo
  {author} {\bibfnamefont{Q.}~\bibnamefont{Yuan}}, \bibinfo {author}
  {\bibfnamefont{Y.-L.~S.}\ \bibnamefont{Tsai}},\ and\ \bibinfo {author}
  {\bibfnamefont{Y.-Z.}\ \bibnamefont{Fan}}}%
   (\bibinfo {year} {2016}),\
  \Eprint{http://arxiv.org/abs/1610.03840}{arXiv:1610.03840 [astro-ph.HE]}%
  \bibAnnoteFile{NoStop}{Cui_2016}%
\end{thebibliography}%

\clearpage

\section{Supplemental Material}
In this Supplemental Material we present a more extensive discussion of the theoretical setup of the analysis,
and we provide some further results and figures to complement those described in the letter.

\section{Theoretical Setup}

The propagation of charged CRs can be described by a diffusion equation \cite{StrongMoskalenko_CR_rewview_2007} for the 
particle density $\psi_i$ of species $i$ per volume and absolute value of momentum $p$
\begin{eqnarray}
  \label{eqn::PropagationEquation}
  \frac{\partial \psi_i (\bm{x}, p, t)}{\partial t} = 
    q_i(\bm{x}, p) &+&  
    \bm{\nabla} \cdot \left(  D_{xx} \bm{\nabla} \psi_i - \bm{V} \psi_i \right) \nonumber \\ 
     +  \frac{\partial}{\partial p} p^2 D_{pp} \frac{\partial}{\partial p} \frac{1}{p^2} \psi_i  & - & 
    \frac{\partial}{\partial p} \left( \frac{\diff p}{\diff t} \psi_i  - \frac{p}{3} (\bm{\nabla \cdot V}) \psi_i \right)  \nonumber   \\ 
    - \frac{1}{\tau_{f,i}} \psi_i - \frac{1}{\tau_{r,i}} \psi_i\,.
\end{eqnarray} 
The source term of primary CRs is denoted by $q_i(\bm{x}, p)$, while the terms proportional to $D_{xx},  \bm{V}$ and $D_{pp}$ correspond 
to CR diffusion, convection and reacceleration, respectively. Furthermore, \eqnref{PropagationEquation} includes the momentum gain or loss rate $\propto {\diff p}/{\diff t}$, adiabatic energy losses $\propto \bm{\nabla \cdot V}$, and loss by fragmentation and radioactive decay $\propto 1/\tau_{f,i}$ and $1/\tau_{r,i}$, respectively. 

Diffusion is modelled by a power law in rigidity $R=p/|Z|$, 
\begin{equation}
  \label{eqn::Dxx}
D_{xx} = D_0 \beta (R/4~{\rm GV})^\delta \,,
\end{equation}
where $\beta = v/c$ denotes the CR velocity. The coefficient of the reacceleration term, $D_{pp}$, is related to $D_{xx}$ and the velocity of Alfven magnetic waves, $v_A$, 
\begin{equation}
  \label{eqn::DiffusivReaccelerationConstant}
  D_{pp} = \frac{4 \left(p \, v_\mathrm{A} \right)^2 }{3(2-\delta)(2+\delta)(4-\delta)\, \delta \, D_{xx}}\,,
\end{equation}
where $\delta$ is the index of the power law as introduced in \eqnref{Dxx}. We assume that convective winds are orthogonal to the Galactic plane, such that $\bm{V}(\bm{x})= {\rm sign}(z)\, v_{0,c} $. 

The source term of primary CRs, $q_i(\bm{x}, p)$, is assumed to factorise into a space- and rigidity dependent part, 
\begin{eqnarray}
  \label{eqn::SourceTerm_1}
  q_i(\bm{x}, p) = q_i(r, z, R) = q_{0,i} \ q_{r,z}(r,z) \, q_R(R)\,,
\end{eqnarray}
where $r,z$ are cylindrical coordinates with respect to the Galactic center. We model the rigidity dependence as a double broken power law with smooth transitions
\begin{eqnarray}
  \label{eqn::SourceTerm_2}
  q_R(R)     &=&   \left( \frac{R}{R_0} \right)^{-\gamma_1}
                  \left( \frac{R_0^{\frac{1}{s}}+R^  {\frac{1}{s}}}
                              {2(R_0)^{\frac{1}{s}}                } \right)^{-s (\gamma_2-\gamma_1)} \nonumber \\[1mm]
           &\times&       \left( \frac{R_1^{\frac{1}{s_1}}+R^  {\frac{1}{s_1}}}
                              { R_1^{\frac{1}{s_1}}                   } \right)^{-s_1(\gamma_3 - \gamma_2)},
\end{eqnarray}
where $R_0$, $R_1$ are the two break positions, $s$, $s_1$ denote the smoothing factors, and $\gamma_i$ ($i=1,2,3$) the slopes
in the three rigidity ranges in between the breaks. The spatial dependence of the source term is parameterized as 
\begin{eqnarray}
  \label{eqn::SourceTerm_3}
  q_{r,z}(r, z)  &= \left( \frac{r}{r_s} \right)^\alpha \exp \left( -\beta \frac{r-r_s}{r_s} \right) 
                                                    \exp \left( -      \frac{|z|  }{z_0} \right),
\end{eqnarray}
with parameters $\alpha = 0.5$, $\beta=1.0$, $r_s=8.5$ kpc, and $z_0=0.2$ kpc.

In the case of antiprotons from DM  annihilation the source term is given  in the main text.
 
Secondary CRs, including in particular antiprotons, are produced through spallation in the interstellar medium (ISM). Their source term is 
calculated from the particle densities of the primary CRs, $\psi_i$, and the corresponding spallation cross-sections $\sigma_{ij}$, 
\begin{eqnarray}
  \label{eqn::SourceTerm_pbar}
  q(\bm{x},p) & = & \sum\limits_{j=\mathrm{H,He}} n_j(\bm{x}) \nonumber\\[1mm]
   &\times& \sum\limits_{i=\mathrm{p,He}} 
    \int \diff p_i \, \frac{\diff \sigma_{ij}(p, p_i)}{\diff p} \beta_i \, c\, \psi_i(\bm{x},p_i),
\end{eqnarray}
where we assume the ISM to be composed of hydrogen and helium, $j=\mathrm{H,He}$, in proportion 1:0.11. 
We use the antiproton production cross-sections from \cite{TanNg_AntiprotonParametrization_1983}, 
but also adopt the cross-sections provided in the more recent study \cite{Mauro_Antiproton_Cross_Section_2014} for comparison. 
Tertiary antiprotons are taken into account with a formula similar to \eqnref{SourceTerm_pbar},
but using as cross-section the total inelastic non-annihilating antiproton cross-section (see \cite{Moskalenko:2001ya} for more details).

We solve \eqnref{PropagationEquation} numerically using  \textsc{Galprop} \cite{Strong:1998fr,Strong:2015zva}, assuming a steady state regime, $\partial\psi_i/\partial t = 0$, with  time-independent  sources $q_i$. 
The equation is solved in Galactic cylindrical coordinates on a three-dimensional grid in $r,z, E_{\rm kin}$.
We use as grid spacing $\Delta r =1$ kpc, $\Delta z=0.2$ kpc and $\Delta \log E_{\rm kin}= \log 1.5$ starting from $E_{\rm kin, min}=1$ MeV.
We verified that the results are stable when using a finer grid.

\begin{figure*}[t!]
  \vspace{-0.6cm}
  \includegraphics[width=0.95\textwidth]{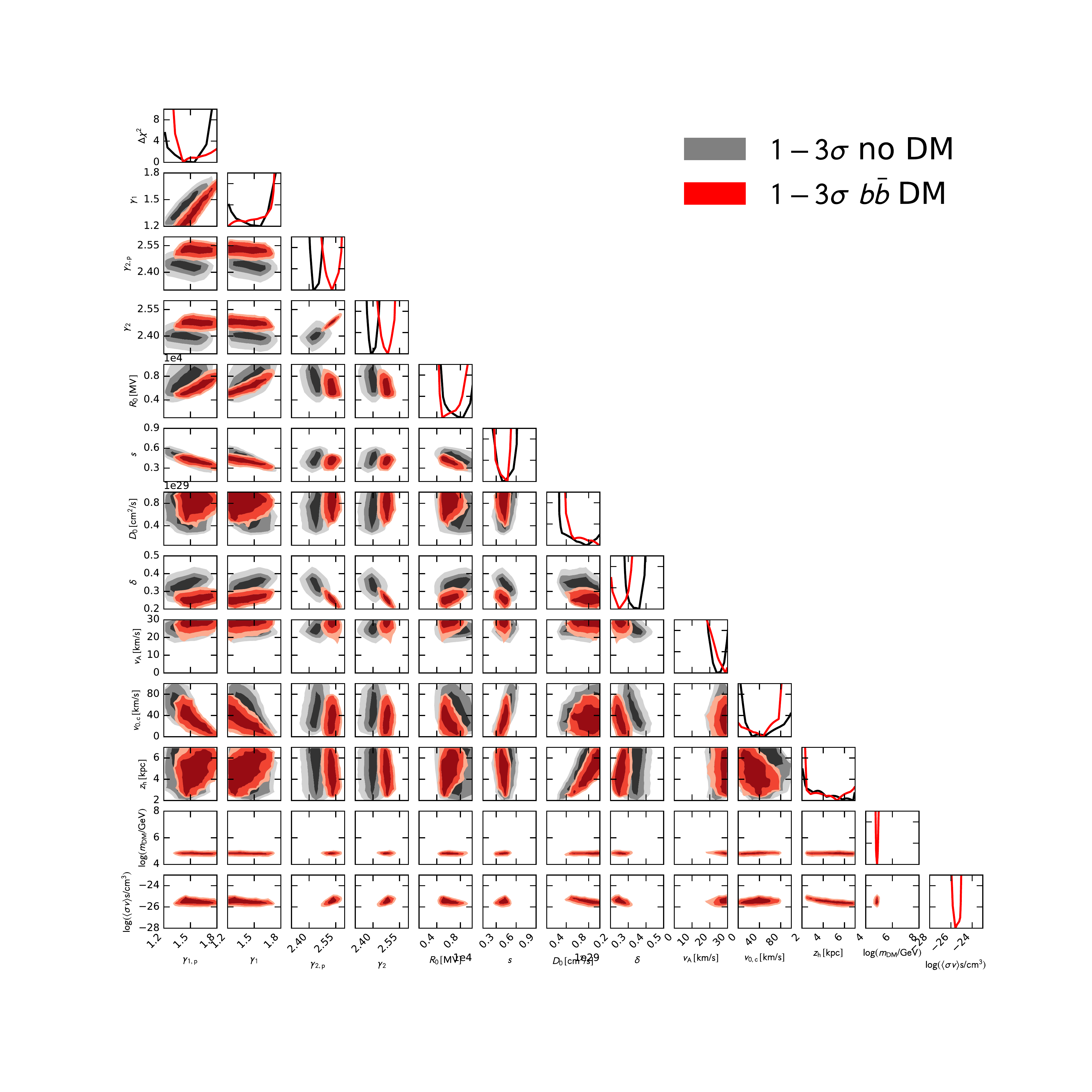}
  \vspace{-0.8cm}
  \caption{Triangle plot for  the cosmic-ray propagation and dark matter fit parameters 
  for the two fits in which DM is included (black contours) or not included (red contours).}
	\label{fig::triangle_full}
\end{figure*}

We use \textsc{MultiNest} \cite{Feroz_MultiNest_2008} to scan  the parameter space defined by \eqnref{PropagationEquation}
and by the DM mass and annihilation cross-section, as summarized in Table I of the main text.
For the \textsc{MultiNest}  settings we use 500 Live Points, an enlargement factor \texttt{efr=0.4} and a tolerance 
\texttt{tol=0.1}. 
The \textsc{MultiNest} scan is effectively thirteen-dimensional, since the three parameters $A_{\rm p}, A_{\rm He}, \phi_{\rm AMS}$
are treated in a special way to exploit the fact that they can be varied without a new \textsc{Galprop} run for given values of 
the other thirteen parameters.  They are, thus, handled as nuisance parameters and profiled away.
In practice, for a   given set of the main'   thirteen parameters the  associated $\chi^2$ is assigned
looking for the minimum $\chi^2$ varying the three nuisance parameters.
For further details on the fit see KC16\,\cite{Korsmeier:2016kha}.

\section{Main fit extended results}

\figref{triangle_full} present the full triangle plot summarizing the results of the main fit with (black contours) and without (red contours) DM.
We can see, as already described in the main text, that the main effect when including DM is a shift of the parameter $\delta$ by about $\sim$30\%.
This shift is accompanied by a corresponding shift in $\gamma_2$ and $\gamma_{2,p}$.
This is expected since the quantity $\delta + \gamma$ has to be equal to the observed slope of the spectra of the
primary species $p$ and He at high rigidities.


\begin{figure*}[!t]
\vspace{-0.5cm}
	\centering
	  \includegraphics[width=8cm] {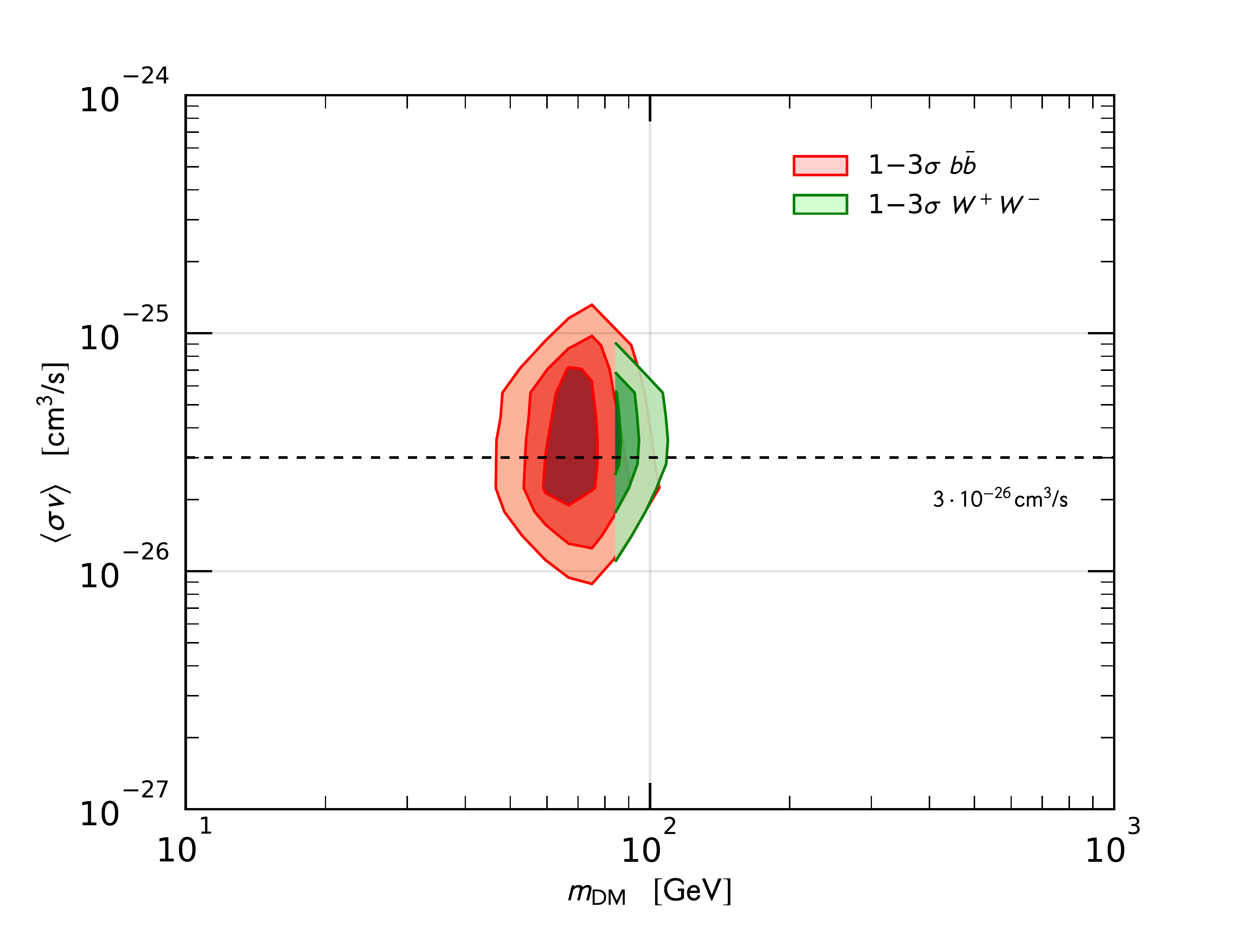} 
          \vspace{-0.5cm}
          \includegraphics[width=8cm] {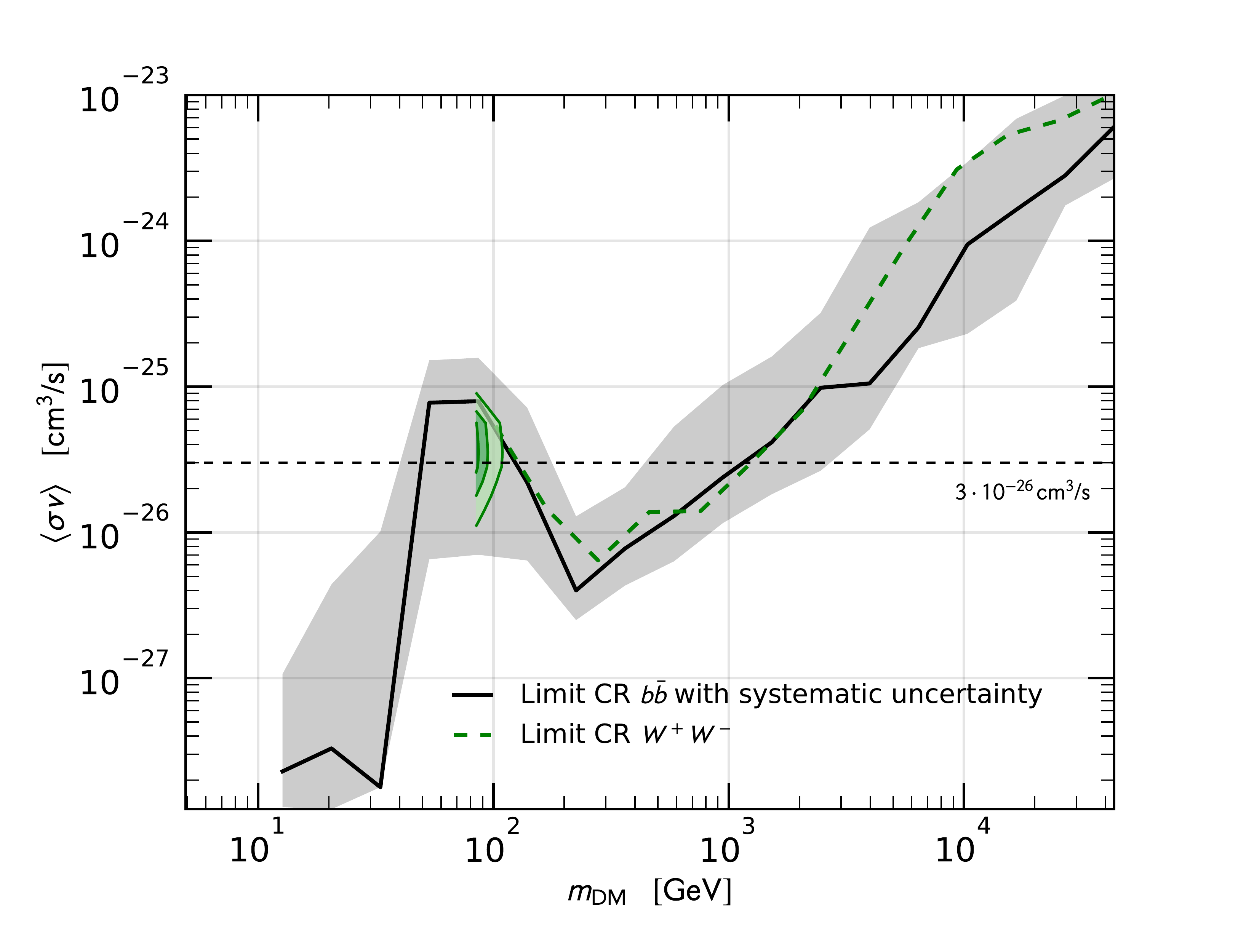}
   	\caption{Left panel: DM best fit regions (1, 2 and 3\,$\sigma$)  for $b\bar{b}$ and $W^+W^-$ final states, respectively. 
	Right panel: limits on the DM annihilation 
	cross-section into $b\bar{b}$ and $W^+W^-$ final states, respectively. See  the text in the main body of the letter for a discussion of the systematic uncertainty represented by the grey shaded band.} 
		  \label{fig::results_WW}
\end{figure*}

\section{Results for $W^+W^-$ annihilation}

In \figref{results_WW} we show the DM preferred region,  and limits on the annihilation cross-section, for $W^+W^-$ final states, 
in comparison to  $b\bar{b}$ final states. 
For $m_{\rm DM} \ge m_W$, where annihilation into $W^+W^-$ is kinematically accessible, the DM preferred region, 
and the limits on the annihilation cross-section, are very similar to those obtained for $b\bar{b}$. 
This is expected since the antiproton spectrum per annihilation is very similar in each hadronic channel,
including $Z^0Z^0$ as well as $u,d,s,c,t$ quarks and gluons.

\section{Comparison with Boron over Carbon}

As discussed in the main text, we fit only light nuclei $p$, $\bar{p}$ and He  to 
take into account the possibility that heavier nuclei could have different propagation properties
and thus bias the result.
Heavier nuclei have different propagation lengths with respect to the light ones, and
therefore probe a different Galaxy volume. As a result, if propagation is non-homogenous,
the fit within a homogenous model would provide inconsistent parameters
if light or heavier nuclei are used. Indeed  indications in this sense have been presented in the literature~\cite{Johannesson_CR_Propagation_2016}.
Nonetheless, the Boron over Carbon ratio has been historically the prime mean to constrain 
propagation of CRs and it is thus useful to check if antiproton propagation
is consistent with Boron and Carbon propagation in the light of the new AMS-02 data.
Here we present some first and still preliminary results and conclusions from such a comparison.
A detailed analysis will be, however, reported in a follow-up publication.

Using the same formalism described for  the fit to $p$, $\bar{p}$ and He,     
we have performed a fit to the recently published B/C AMS02 data \cite{Aguilar:2016vqr}
together with $p$ and He AMS-02 data. Propagation is thus constrained by B/C, while
the joint fit to  $p$ and He data ensures a reliable prediction for the antiproton flux.
The result of the fit to B/C, down to 5 GV is shown in the left panel of \figref{results_BC}.
It can be seen that we are able to achieve a good fit with very flat residuals.
The right panel shows the predicted antiproton over proton ratio from this fit.
The uncertainty band has been derived from a propagation of the errors from the B/C fit. 
No fit to the antiproton flux is performed.
From the residuals panel it can be seen that  the agreement is quite good above.
Notably, the same excess feature at about 20 GV present in the $\bar{p}/p$ fit also appear here.
The parameters of the fit for the two  cases are indeed very close, except for a slightly larger value
of $\delta$ ($\approx$ 0.35 in this case, vs $\approx$ 0.25 for the $\bar{p}/p$ fit).

This preliminary result thus indicates that light and heavier nuclei have compatible propagation, 
and that an homogenous diffusion scenario is still in agreement with the data.
A joint fit with $\bar{p}/p$ and B/C and DM should thus be possible (see also \cite{Cui_2016,Yuan:2017ozr,Jin:2017iwg,Lin:2016ezz}), and would provide 
more stringent constraints. This will be explored in follow-up analyses.

\begin{figure*}[!t]
\vspace{-0.5cm}
	\centering
	  \includegraphics[width=8cm] {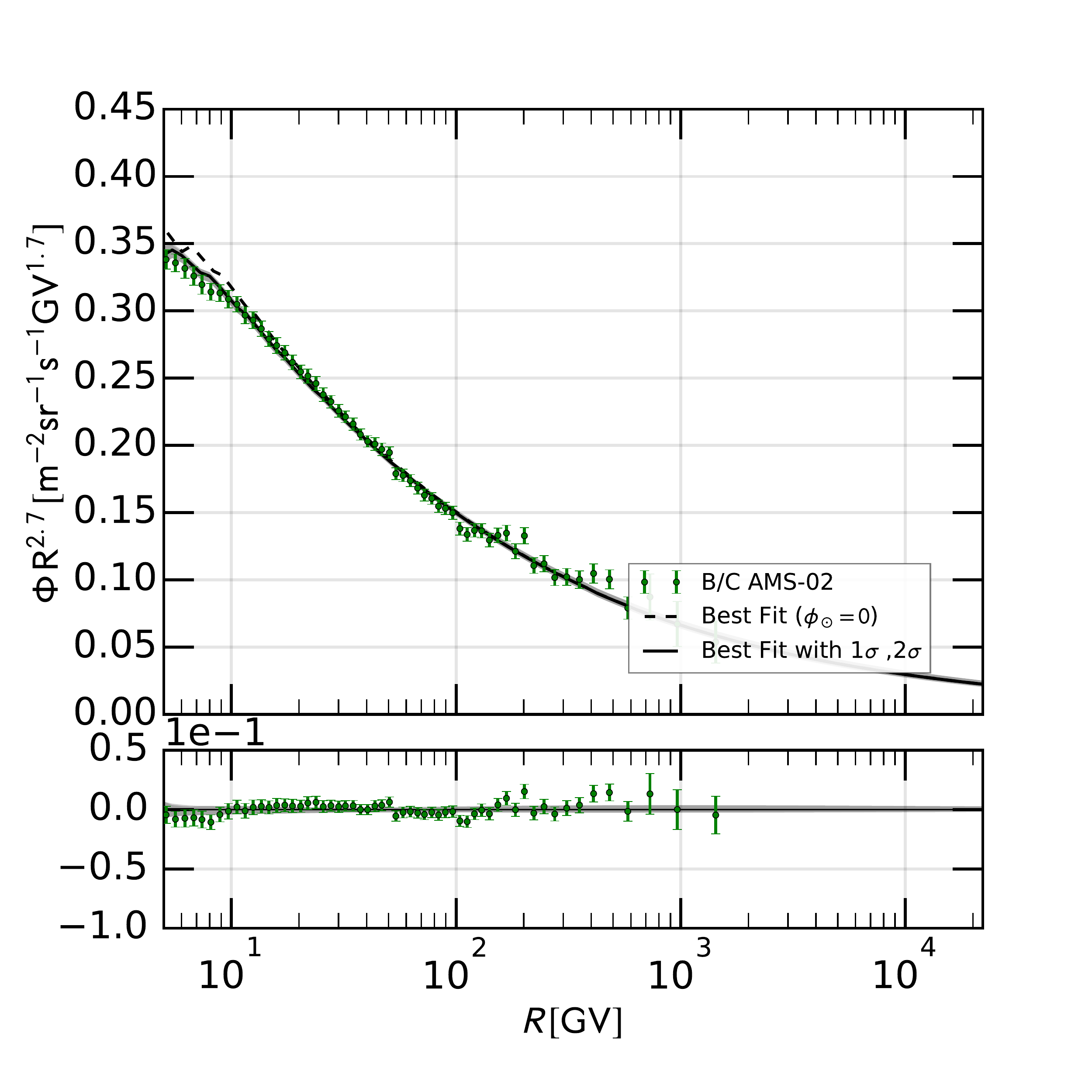} 
          \vspace{-0.5cm}
          \includegraphics[width=8cm] {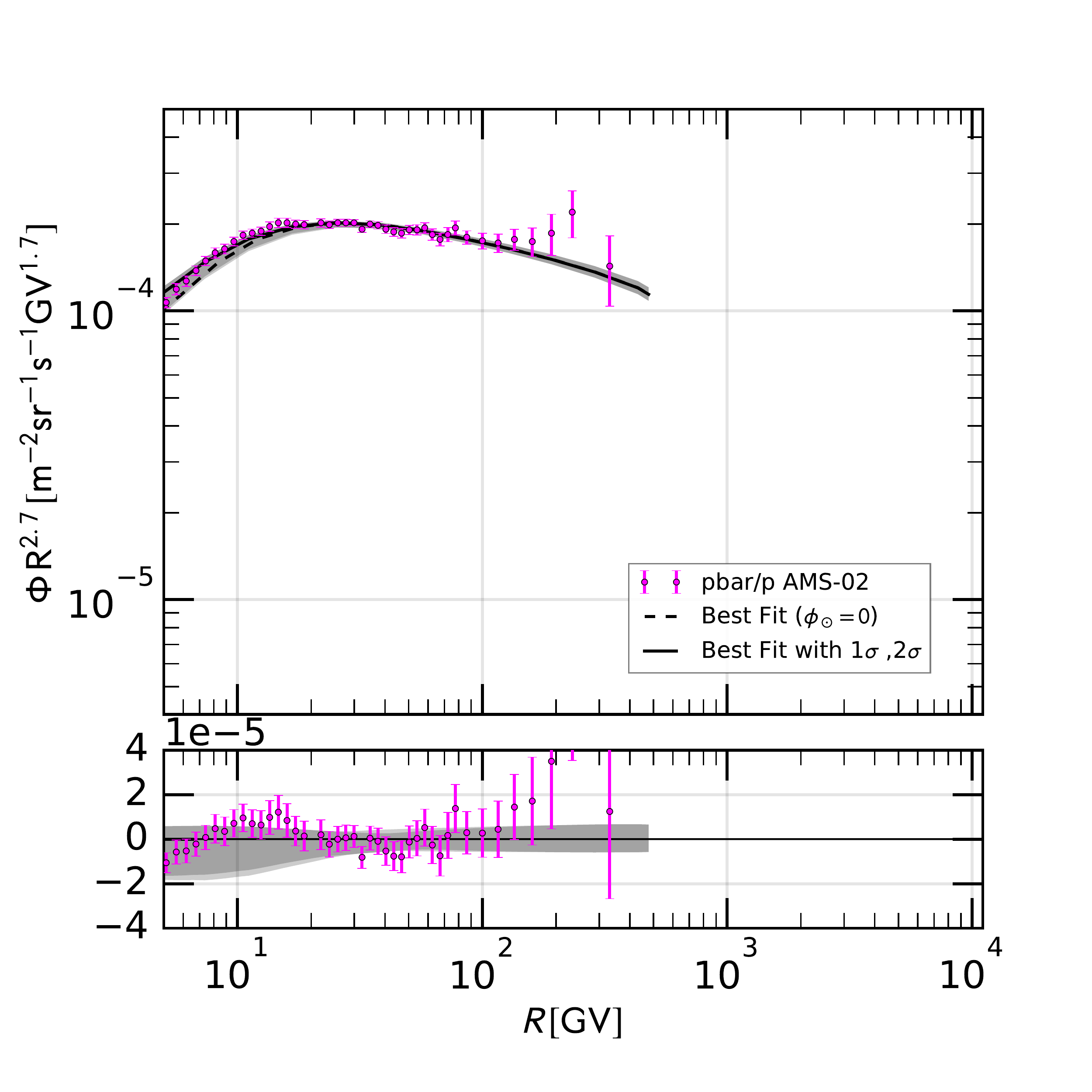}
   	\caption{Left panel: Best fit and residuals for the Boron over Carbon data. 
	Right panel: predicted antiproton over proton ratio from the B/C fit, and residuals.} 
		  \label{fig::results_BC}
\end{figure*}

\end{document}